\documentclass{aa}  

\usepackage{graphicx}
\usepackage{txfonts}
\usepackage{xcolor}
\usepackage{placeins}
\usepackage{float}
\usepackage{caption}
\usepackage{multirow}
\begin{document}

   \title{Submillimeter galaxy overdensities around physically associated quasar pairs}

   \author{Eileen Herwig\inst{1}
        \and
        Fabrizio Arrigoni Battaia\inst{1}
        \and
        Chian-Chou Chen\inst{2}
        \and
        Aura Obreja\inst{3,4}
        \and
        Marta Nowotka\inst{5,6}
        \and
        Rhea-Silvia Remus\inst{7}
        \and
        Hidenobu Yajima\inst{8}
        }

   \institute{Max-Planck-Institut f\"ur Astrophysik, Karl-Schwarzschild-Straße 1, D-85748 Garching bei M\"unchen, Germany\\
              \email{eherwig@mpa-garching.mpg.de}
              \and
              Academia Sinica Institute of Astronomy and Astrophysics (ASIAA), 11F of Astronomy-Mathematics Building, AS/NTU, No. 1, Section 4, 12 Roosevelt Road, Taipei 106319, Taiwan
              \and
              Interdisciplinary Center for Scientific Computing (IWR), University of Heidelberg, Im Neuenheimer Feld 205, D-69120 Heidelberg, Germany
              \and
              Universit\"at Heidelberg, Zentrum für Astronomie, Institut f\"ur Theoretische Astrophysik, Albert-Ueberle-Straße 2, D-69120 Heidelberg, Germany
              \and
              Department of Physics, Stanford University, Stanford, CA 94305, USA
              \and
              Kavli Institute for Particle Astrophysics \& Cosmology, P.O. Box 2450, Stanford University, Stanford, CA 94305, USA
              \and
              Universit\"ats-Sternwarte, Fakult\"at f\"ur Physik, Ludwig-Maximilians-Universit\"at Mnchen, Scheinerstr.~1, 81679 M\"unchen, Germany
              \and
              Center for Computational Sciences, University of Tsukuba, 1-1-1 Tennodai, Tsukuba, Ibaraki 305-8577, Japan
             }

   \date{}
 
  \abstract{A commonly employed method to detect protoclusters in the young universe is the search for overdensities of massive star forming galaxies, such as submillimeter galaxies (SMGs), around high-mass halos, including those hosting quasars. In this work, we study the Megaparsec environment surrounding nine physically associated quasar pairs between $z=2.45$ and $z=3.82$ with JCMT/SCUBA-2 observations at 450~$\mu$m and 850~$\mu$m covering a field of view of roughly 13.7$\arcmin$ in diameter (or 32~Mpc$^2$ at the median redshift) for each system. We identify a total of 170 SMG candidates and 26 non-SMG and interloper candidates. 
  A comparison of the underlying 850~$\mu$m source models recovered with Monte Carlo simulations to the blank field model reveals galaxy overdensities in all fields, with a weighted average overdensity factor of $\delta_{\rm cumul} = 3.4 \pm 0.3$.
  From this excess emission at 850~$\mu$m, we calculate a star formation rate density of $1700 \pm 100$~M$_{\odot}$~yr$^{-1}$~Mpc$^{-3}$, consistent with predictions from protocluster simulations and observations. Compared to fields around single quasars, those surrounding quasar pairs have higher excess counts and more centrally peaked star formation, further highlighting the co-evolution of SMGs and quasars. We do not find preferential alignment of the SMGs with the quasar pair direction or their associated Ly$\alpha$ nebulae, indicating that cosmic web filaments on different scales might be traced by the different directions. Overall, this work substantiates the reliability of quasar pairs to detect overdensities of massive galaxies and likely sites of protocluster formation. Future spectroscopic follow-up observations are needed to confirm membership of the SMG candidates with the physically associated quasar pairs and definitively identify the targeted fields as protoclusters.
  }

   \keywords{ Submillimeter: galaxies - Galaxies: evolution - Galaxies: interactions - Galaxies: high-redshift - Galaxies: halos – quasars: general - (Cosmology:) large-
scale structure of Universe
               }

   \maketitle
%

\section{Introduction}

Quasars, highly accreting active galactic nuclei (AGNs), are among the most optically luminous types of galaxies and thus detectable out to large distances. With their estimated halo masses of $10^{12} - 10^{13}$ M$_{\odot}$ \citep{Shen2007, White2012, Eftek2015, Costa2023}, they are among the most massive halos before and around cosmic noon \citep{Behroozi2013, Hernandez-Aguayo2022}. Thus, they have become an invaluable tool in pinpointing the highest density peaks in the young universe (e.g. \citealt{Falder2011, Garcia-Vergara2019, Fossati2021, Luo2022, FAB2023, Pudoka2024, Champagne2025}), similarly to the even rarer high-redshift radio galaxies (e.g. \citealt{Venemans2007, Miley2008, Orsi2016}).
Following hierarchical structure formation \citep{Springel2005}, such overdensities are the most likely birthplaces of galaxy clusters, the most massive virialized structures in the local universe reaching halo masses between $10^{14}$ and $10^{15}$~M$_{\odot}$ at $z=0$ \citep{Overzier2016}. At cosmic noon, their progenitors, the so-called galaxy protoclusters, consist of a large number of mainly star-forming galaxies spread across a vast area of multiple megaparsecs (Mpc). This environment often hosts a massive, central galaxy with exceptional activity, such as a quasar or a radio-loud AGN. This galaxy eventually evolves into the brightest cluster galaxy through subsequent mergers \citep{Muldrew2015, Chiang2017, Lovell2018, Yajima2022, Rennehan2024}. 

Due to the relative large number of galaxies in these structures, interactions are hypothesized to trigger dust-obscured star-formation in the more massive protocluster members. Bright emission from the heated dust leads to detections of such sources as submillimeter galaxies (SMGs) at wavelengths between a few hundred and a thousand microns, but they remain optically faint because of dust obscuration \citep{Smail1997, Barger1998, Engel2010, Vieira2010, Casey2014, Smail2014, Hung2016, Oteo2018}. Mainly found at $z=1-3$ \citep{Barger2000, Chapman2005, Wardlow2011, Danielson2017, Dudzeviciut2020, TC2022}, SMGs are extraordinary sites of hidden starbursts in massive galaxies: they exhibit estimated star formation rates (SFRs) of up to multiple thousand solar masses a year \citep{Riechers2013, Swinbank2014, Tacconi2018, Quiros-Rojas2024}, and populate the massive end of the stellar mass function at their cosmic epoch with stellar mass estimates between $10^{10}$ and $10^{11}$~M$_{\odot}$ \citep{Michalowski2012, daCunha2015ALESS, Michalowski2017,Dudzeviciut2020, Liao2024, Gottumukkala2024}.

In line with expectations, submillimeter studies of the Mpc-scale environment around different types of active-galactic-nuclei (AGN) have revealed -- in some cases -- increased number counts of SMGs compared to the blank field \citep{Stevens2003, Humphrey2011, Rigby2014, Dannerbauer2014, Jones2017, Wethers2020, FAB2023, Zhou2024, Wang2025}. Nevertheless, AGNs remain controversial as tracers of protoclusters as they do not always pinpoint galaxy overdensities \citep{Uchiyama2018, Champagne2018, Cornish2024}. However, recent works targeting the fields of unobscured bright quasars with the Submillimetre Common-User Bolometer Array 2 (SCUBA-2) instrument at the James Clerk Maxwell Telescope (JCMT) found ubiquitous excess counts of SMGs around 
single bright quasars and multiple physically associated quasars, with the latter showing higher overdensity factors (\citealt{Nowotka2022, FAB2023}). Given the small number of multiple quasar fields targeted (only four), those studies were not conclusive in assessing whether multiple quasars (from quasar pairs onwards) are a more reliable tracer of protocluster environments as usually suggested \citep{Onoue2018, Sandrinelli2018}.

\begin{table*}
\caption{Information on the quasar pair at the center of the nine observed fields, the coordinates of the pointing and observing conditions.}
\label{tab:obslog}
\centering 
\begin{tabular}{c c c c c c c c c}
\hline\hline
ID\tablefootmark{a} & Quasar\tablefootmark{b} & RA (J2000)\tablefootmark{c} & Dec (J2000)\tablefootmark{c} & $z$ & $d$ [kpc/\arcsec]\tablefootmark{d} & \#Scans & $\tau$ & $A_{\rm eff}$ [\arcmin$^2$]\\    
\hline
\multicolumn{8}{c}{In QSO MUSEUM II \citep{Herwig2024}} \\ \hline
1 & QSO B2359+068 & 00:01:40.59 & +07:09:47.81  & 3.23 & 48/6.2 & 5 & 0.04-0.05 & 102.2\\
2 & SDSS J001807.36+161257.5 & 00:18:08.09 & +16:12:50.84  & 3.13 & 97/12.5 & 3 & 0.07 & 100.7\\
3 & SDSS J024005.23-003909.8 & 02:40:05.23 & -00:39:09.84  & 3.12 & 69/8.8 & 6 & 0.03-0.08 & 108.7\\
4 & SDSS J024442.58-002320.4 & 02:44:42.60 & -00:23:20.40  & 3.04 & 482/61.4 & 3 & 0.04-0.05 & 106.0\\
5 & SDSS J101254.73+033548.7 & 10:12:54.74 & +03:35:48.84  & 3.16 & 478/61.6 & 6 & 0.02-0.03 & 107.4\\
6 & SDSS J102116.98+111227.6 & 10:21:16.47 & +11:12:27.94  & 3.82 & 55/7.6 & 6 & 0.02-0.1 & 106.9 \\ \hline
\multicolumn{8}{c}{Other} \\ \hline
ELAN0101 & QSO J0101+0201 & 01:01:16.54 & +02:01:57.4 & 2.45 & 75/9.0 & 6 & 0.04-0.09 & 103.0\\
J0119 & SDSS J011907.46+020558.2 & 01:19:07.46 & +02:05:58.19  & 3.00 & 435/55.2 & 4 & 0.02-0.06 & 104.8\\
J1622 & 2MASS J16221011+0702153 & 16:22:09.81 & +07:02:11.54  & 3.26 & 45/5.8 & 5 & 0.01-0.03 & 110.0\\

\hline
\end{tabular}
\tablefoot{
\tablefoottext{a}{Matched to the IDs assigned in QSO MUSEUM II \citep{Herwig2024}.}
\tablefoottext{b}{Identifier of the brighter quasar in the pair.}
\tablefoottext{c}{Coordinates of the telescope pointing.}
\tablefoottext{d}{Projected distance between the quasar pair.}
}
\end{table*}

In the presence of large numbers of detected SMGs, one can start to study their distribution and properties with respect to the distance to, and structures associated with the central active massive galaxy.
On average, the SMGs in overdense structures around radio galaxies have been found to align in a preferential direction roughly perpendicular to the radio-jet axis \citep{Zeballos2018}. This distribution, also seen around individual radio galaxies (e.g., \citealt{Zhou2024}), has been interpreted as a large-scale filament feeding the assembling cluster \citep{Zeballos2018}. 
Similarly, \cite{FAB2023} found that the SMGs around bright quasars define preferred directions, which can be used to stack the individual fields and uncover an elongated (tens of cMpc) structure reminiscent of a large-scale filament with a scale width of $\approx3$~cMpc. The same study shows that, on average, the direction of each SMG distribution roughly aligns with the major axis of the extended ($\sim 100$~kpc; e.g. \citealt{FAB2019a}) Ly$\alpha$ emission around the central quasar.
Such emission is believed to originate from cool ($10^4$~K) gas reservoirs in the circumgalactic medium of the quasars and thus likely associated with gas accretion from cosmic web filaments \citep{FAB2018a, Costa22}. 
This idea is corroborated by the finding that Ly$\alpha$ emission around physically associated quasar pairs exhibits alignment with the direction of the pair \citep{Herwig2024,Tornotti2025}.
However, due to the high halo mass of quasars, the extended Ly$\alpha$ around quasar pairs likely traces a short (up to $\sim1$~ physical Mpc) filament connecting the close-by galaxies \citep{Tornotti2025}.

In order to provide further insight into the relation between the number of central quasars and the richness of their environment, 
this paper investigates the Mpc-scale galaxy environment of nine additional quasar pairs at submillimeter wavelengths.
In contrast to previous works, the majority of our sample presents only modestly extended Ly$\alpha$ emission \citep{Herwig2024}, with the addition of one known Enormous Ly$\alpha$ nebula (ELAN) field \citep{Cai2018}. 
The rare ELANe can reach projected distances of up to 500~kpc from the central source, far exceeding the quasar's expected circumgalactic medium (100~kpc), and are characterized by integrated nebula luminosities above $10^{44}$~erg~s$^{-1}$ \citep{Cantalupo2014, Hennawi2015, Cai2017, FAB2018a, Cai2018, Li2024}. They are usually associated with two quasars, though the first known high-$z$ quadruple quasar is also embedded into an ELAN, and the over-abundance of cool hydrogen gas implied by the Ly$\alpha$ emission has been interpreted as fuel for the assembly of a massive galaxy cluster \citep{Hennawi2015}.
Also, the sample includes a newly proposed physically-associated quadruple quasar candidate \citep{Herwig2025}, with an extreme overdensity of at least three massive galaxies within only 20~kpc in projection.

The paper is organized as follows. We describe our observations and data reductions in Sec.~\ref{sec:obs}, detail the source extraction and source model reconstruction in Sec.~\ref{sec:extraction} and Sec.~\ref{sec:diffnum}, and present the resulting galaxy overdensities, alignments and SFRs in Sec.~\ref{sec:results}. We discuss our results in Sec.~\ref{sec:discussion} and summarize this work in Sec.~\ref{sec:summary}.
Throughout this paper, we assume a flat ${\Lambda}$CDM cosmology with ${H_0\ =\ 67.7}$~km~s$^{-1}$~Mpc$^{-1}$, ${\Omega_m\ =\ 0.31}$, and ${\Omega_{\Lambda}\ =\ 0.69}$ \citep{Planck18cosmo}.

\section{Observations and data reduction}
\label{sec:obs}

\begin{figure}
    \centering
    \includegraphics[width=\hsize]{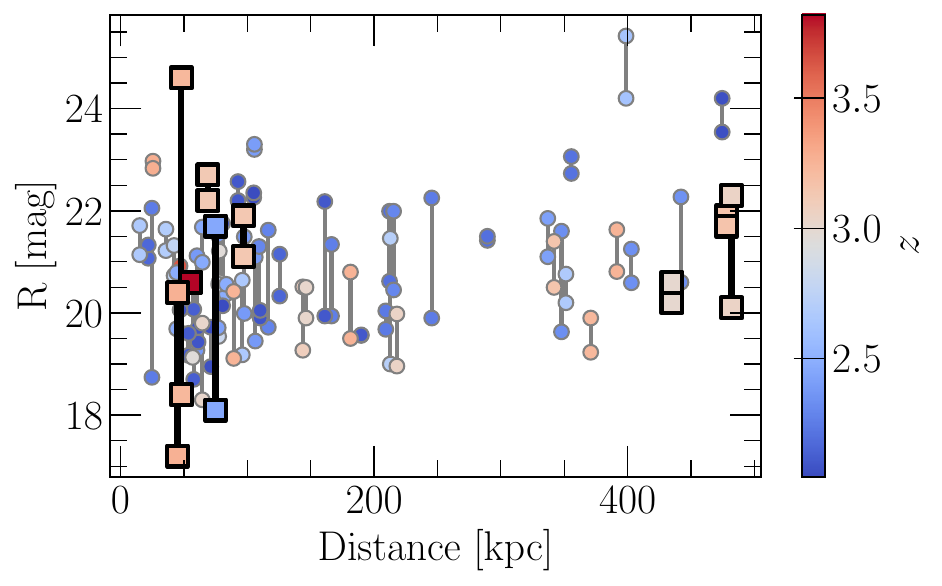}
    \caption{Magnitude and projected distance of the quasar pairs presented in this work (squares) and found in catalogs (circles). The points are color-coded by redshift.}
    \label{fig:sample}
\end{figure}

   \begin{figure*}[h]
   \centering         
   \includegraphics[width=0.87\hsize]{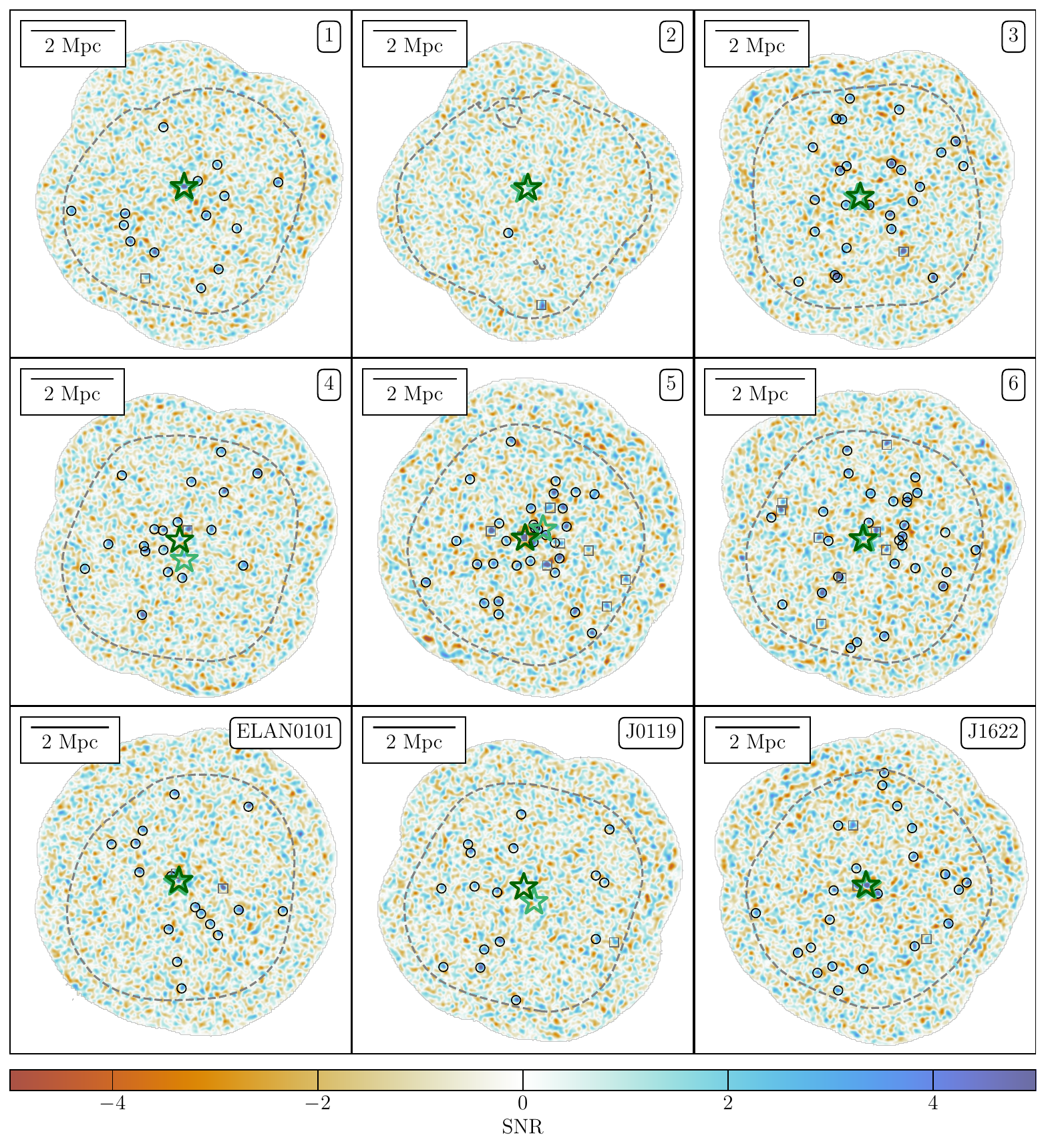}
      \caption{SNR maps at 850~$\mu$m. The position of the brighter (fainter) quasar is indicated by a darkgreen (lightgreen) star. Identified catalog SMGs are marked by black circles. The grey dashed line indicates the edge of the effective area (2.5 $\times$ the central rms). Sources inconsistent with the expected SMG SED at the quasar redshift are marked with grey rectangles.
              }
         \label{fig:snrmaps}
   \end{figure*}

We rely on submillimeter data of nine quasar pair fields obtained with JCMT/SCUBA-2 \citep{Holland2013}) under the program IDs M22AP049 and M22BP025. Information on the targeted fields and the observing conditions is summarized in Table~\ref{tab:obslog}, and Table~\ref{tab:quasars} shows positions and magnitudes of the quasars in the pairs.

The sample is selected from the SDSS DR12 quasar catalog \citep{Paris2017} and AGN searches at high-$z$ (e.g., \citealt{Bielby2013}) as described in \cite{FAB2019b}, \cite{Cai2018} and \cite{Herwig2024}. Specifically, quasars are required to be within $1 \arcmin$ (or approximately 500~kpc at the median redshift) and $\Delta z \leq 0.03$ (or roughly $\Delta v < 2000$~km~s$^{-1}$) of each other to achieve a high likelihood of physical association and therefore a local AGN overdensity. The observed sample spans a redshift range between $z = 2.45$ and $z = 3.83$ with a median redshift of 3.13.
We visualize magnitude, projected pair separation and redshift of our sample compared to pairs found in the Million Quasars (Milliquas) v8 catalog \citep{Flesch2023} and the aforementioned catalogs between $z=2$ and $z=4$ and on the full sky. While the magnitudes of our sample are typical, the observed pairs do not cover intermediate projected separations between 150~kpc and 400~kpc.

For seven pairs or pair members, rest-frame ultraviolet integral-field unit observations are available probing Ly$\alpha$ emission on scales of the circumgalactic medium within a few hundred kiloparsecs around the quasars using the Multi-Unit Spectroscopic Explorer on the Very Large Telescope (IDs 1 to 6, \citealt{Herwig2024}) or the Palomar and Keck Cosmic Web Imager (ELAN0101, \citealt{Cai2018}).
The SCUBA-2 observations were performed with an exposure time of about 30 minutes per scan between May 24, 2022 and January 11, 2023 under very good ($\tau < 0.05$, program ID M22BP025) and good ($\tau < 0.08$, program ID M22AP049) weather conditions using a Daisy scanning pattern centered on one of the quasars in the pair and covering a field of view (FoV) of around $13.7 \arcmin$ in diameter. At the median redshift of the sample, this corresponds to a diameter of 6.4~physical Mpc and is comparable to the expected size of protoclusters at these redshifts \citep{Chiang2017}.
We obtained observations of the 450~$\mu$m band with an effective beam FWHM of 9.8\arcsec\ and of the 850~$\mu$m band with an effective beam FWHM of 14.6\arcsec.

We performed the data reduction following the methods described in \cite{TC2013a} and employed in \cite{Nowotka2022} and \cite{FAB2023}, utilizing the Dynamic Iterative Map Maker (DIMM) from the SMURF package within the STARLINK software \citep{Jenness2011, Chapin2013}. We processed each scan using the standard configuration file, dimmconfig\_blank\_field.lis. The reduced scans were then combined into final maps using the MOSAIC\_JCMT\_IMAGES recipe in PICARD, the Pipeline for Combining and Analyzing Reduced Data \citep{Jenness2008}. To enhance point source detectability, we applied a standard matched filter to the final maps using the PICARD recipe SCUBA2\_MATCHED\_FILTER. Finally, we employed the latest flux conversion factors of 472~Jy~beam$^{-1}$~pW$^{-1}$ for the 450~$\mu$m map and 495~Jy~beam$^{-1}$~pW$^{-1}$ at 850~$\mu$m \citep{Dempsey2013, Mairs2021}.
In the 850~$\mu$m maps of ID2, two out of three exposures show interference patterns and we excluded them from the combination and analysis.

This procedure results in one signal map and one noise map per observed field and wavelength band. Excluding ID2, the central noise in the final maps spans from 6.54~mJy~beam$^{-1}$ to 13.94~mJy~beam$^{-1}$ at 450~$\mu$m and from 0.79~mJy~beam$^{-1}$ to 1.24~mJy~beam$^{-1}$ at 850~$\mu$m. The noise increases significantly towards the edges of the maps and thus we only considered pixels below 2.5 times the central noise for the following analysis. The region satisfying this condition is called the effective area $A_{\rm eff}$.

To obtain true noise maps, we employed the jackknife technique by dividing the exposures for each band and field into two maps containing roughly half of the exposure time each, and subtracting these maps from each other. We rescaled the two maps by the factor $\sqrt{t_1 \times t_2}/(t_1 + t_2)$ to match the exposure times $t_1$ and $t_2$ in the two maps. Because of the strong weather dependence of the data sensitivity at 450~$\mu$m, we ensured that the weather conditions in the exposures of each image are roughly equal.
In this way, any true source irrespective of its significance will be removed from the jackknife map and only noise features remain.
As only one exposure at 850~$\mu$m for the field of quasar pair 2 is available, we cannot construct jackknife maps for the object and exclude it from any further analysis requiring true noise maps. For quasar pair 4, only three exposures are available and we cannot divide them according to weather conditions, leading to an increased central noise of 14.8~mJy~beam$^{-1}$ at 450~$\mu$m.
The central noise in the rest of the jackknife maps is consistent with the central value in the noise maps, deviating only by a maximum of 2.9~\% at 850~$\mu$m, with values between 0.78~mJy~beam$^{-1}$ and 1.28~mJy~beam$^{-1}$, and up to 6.8~\% at 450~$\mu$m, spanning from 6.50~mJy~beam$^{-1}$ to 12.1~mJy~beam$^{-1}$.

\section{Source extraction}
\label{sec:extraction}

We extracted sources that lie fully within the respective effective area by identifying the highest SNR peak, writing relevant information to the source catalog and removing this source from the map by subtracting a scaled point spread function (PSF) as galaxies are not resolved at the low spatial resolution of the data. We repeated this process until the highest SNR value fell below three. Sources consistent (i.e. within 6\arcsec) with any quasar positions in these initial source catalogs and their counterpart source detection in the other band are summarized in Table~\ref{tab:quasars} of Appendix~\ref{app:quasars}.

To increase the reliability of source detection, we cleaned the catalogs by removing any source below 4$\sigma$ that is not detected in both bands above 3$\sigma$. Further, we identified all sources from the catalog at 850~$\mu$m that are detected in both bands, but whose deboosted fluxes within errors do not fit the ratio expected from an SMG SED \citep{daCunha2015ALESS} at the redshift of the quasar pair and $\pm$~9,000~km~s$^{-1}$ away from it. These are likely low redshift interlopers, multiple blended sources or SMGs with significant contribution from a dust-obscured AGN, leading to an atypical SED. In total, we identified 26 interloper or non-SMG candidates, with a median value of 1 per field, and summarized them in Table~\ref{tab:interlopers} within Appendix~\ref{app:interloper}.

In this way, we obtained two source catalogs for each quasar pair field with more detections at 850~$\mu$m (median 20 excluding ID2) than at 450~$\mu$m (median 3) due to the lower noise level in the first band. In total, we identified 170 SMG candidates in the 850~$\mu$m band (Fig.~\ref{fig:snrmaps}, Table~\ref{tab:smg1}-\ref{tab:smg1622}).

\subsection{Reliability}
\label{subsec:reliab}

A commonly adopted method for determining the reliability of source extraction is the inversion of the signal map and extraction of a source catalog from the inverted map. This method relies on the assumption that noise peaks have the same absolute value as noise troughs (i.e. peaks in the inverted map). However, this is not the case for maps processed with the matched filter approach as it introduces very negative values around the PSFs of the brightest sources. Thus, in median, we detected eight spurious sources in the inverted 850~$\mu$m maps (excluding ID2), which show more numerous and more significant source detections, but only a median of one spurious source at 450~$\mu$m.

We therefore determined the reliability of our source detection using the jackknife maps instead, as done in \cite{FAB2023}. For this, we extracted sources from the jackknife map following the same procedure as detailed in Sec.~\ref{sec:extraction}. Since spurious sources do not have counterparts, we only extracted data peaks above 4$\sigma$. We found a median number of 1.5 spurious sources at 450~$\mu$m and 0.5 spurious sources at 850~$\mu$m, in agreement with or lower than previous works \citep{FAB2018, FAB2023}. As a similar number of false detections is expected to occur in the source catalogs at 850~$\mu$m, we obtain a reliability of 97.6~\%.

\subsection{Completeness}

\begin{figure}[h]
   \centering         
   \includegraphics[width=\hsize]{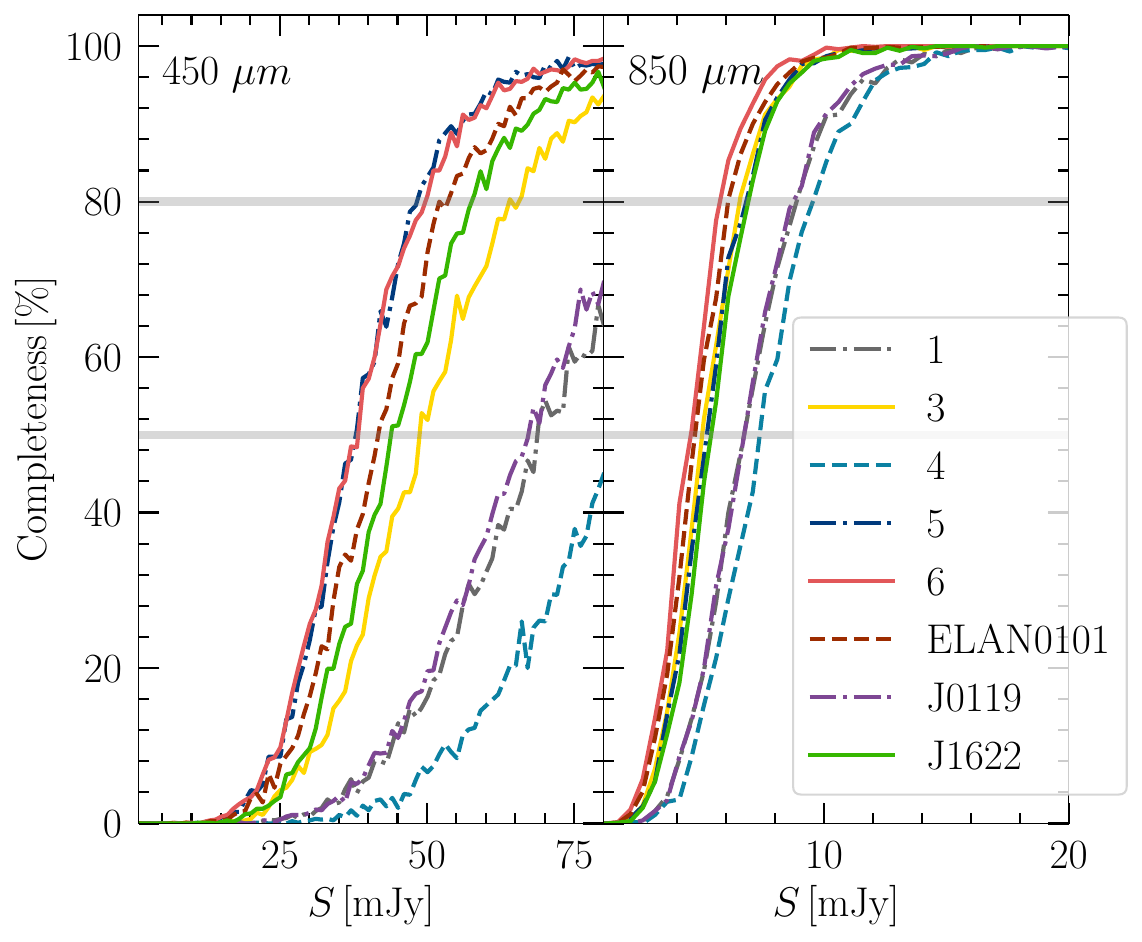}
      \caption{Completeness percentage of the source extraction as a function of injected source flux in the 450~$\mu$m band (left) and the 850~$\mu$m band (right).
              }
         \label{fig:completeness}
   \end{figure}

We determined the flux at which the source extraction can be considered complete by injecting sources into the effective area of the jackknife maps using the same PSF as during source extraction and attempting to recover them using the same algorithm as in Section~\ref{sec:extraction}. In the map of the 850~$\mu$m band, we injected 10,000 sources each for fluxes between 0.1~mJy and 30.1~mJy in 0.5~mJy steps, while at 450~$\mu$m, we used fluxes between 0.1~mJy and 80.1~mJy in 1~mJy steps. Fig.~\ref{fig:completeness} displays the resulting completeness percentage per flux.

The completeness is mainly affected by the weather conditions during the observations and the number of scans used in the combined map and can stay particularly low at 450~$\mu$m ($< 50$~\% in field~4 for any injected source flux).

\section{Differential number counts}
\label{sec:diffnum}

\begin{figure*}[h]
    \centering
    \includegraphics[width=0.87\hsize]{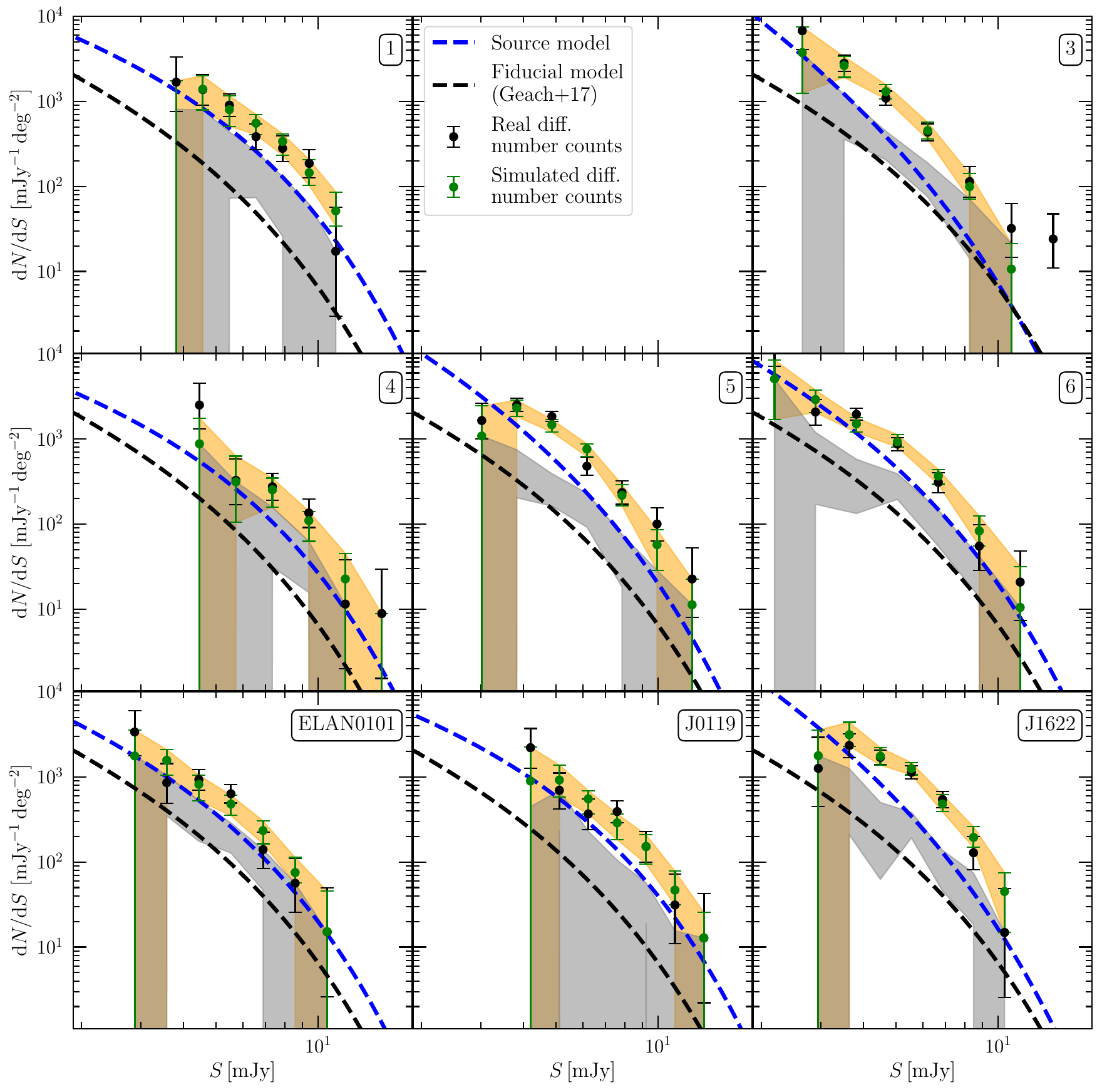}
    \caption{Differential number counts at 850~$\mu$m in the data (black points) and the simulated maps (green points). The underlying source model of the quasar pair fields (blue dashed) is shown in comparison to the field model (black dashed) with the 1$\sigma$ spread of the differential number counts from 500 realizations of the source model (yellow shaded area) and the fiducial model (grey shaded area).}
    \label{fig:diffnums}
\end{figure*}
We aim to characterize the SMG overdensity around quasar pairs in comparison to the field and thus require knowledge of the underlying source model in the observed 850~$\mu$m maps centered on quasar pairs, which we reconstructed from differential number counts. We followed the methods described in \cite{TC2013a}.

As a first step, we determined the SNR at which the number of pixels in the data map is significantly larger than in the jackknife map (i.e. $ \geq 3 \times$), meaning that those pixels are mostly dominated by sources.
On average, this condition is fulfilled at a SNR of 2.9, ranging between 2.5 and 3.3. We used these values as a new detection threshold to extract sources from the data maps. Naturally, the lower SNR requirement leads to a lower reliability of source extraction, as quantified by extracting all spurious sources down to the same SNR threshold in the jackknife maps (Sec.~\ref{subsec:reliab}). However, as we are only interested in the raw number counts and fluxes instead of the position of sources, we can use this number to correct the raw number counts. For that, we divided detections into equally spaced logarithmic flux bins and normalized the number counts by bin width and detectable area, defined as the area of the map in which the noise is low enough that a source of the given flux would reach the SNR threshold. After processing the number counts from the jackknife maps in the same way, we subtract these spurious sources from the data number counts.

The resulting data points are affected by multiple observational biases such as flux boosting, source blending and incompleteness, that have to be taken into account to obtain the underlying source model. To reconstruct it by finding a model that reproduces the observed raw number counts, we used Monte Carlo simulations following the exact procedure of \cite{Nowotka2022} that we briefly summarize below. 
For easier comparison with blank field models such as \cite{Geach2017}, we parameterized the model using the same Schechter function as in this work,
\begin{displaymath}
    \frac{dN}{dS} = \frac{N_0}{S_0} \left( \frac{S}{S_0} \right)^{-\gamma} {\rm exp} \left( -\frac{S}{S_0}\right),
\end{displaymath}
with the normalization $N_0$ (in deg$^{-2}$) and the characteristic flux $S_0$ (in mJy), which is fixed to a value of 2.5~mJy to avoid the degeneracy between $N_0$ and $S_0$.
To get a set of initial parameters, we fit the raw number counts in the quasar pair fields with the Schechter function and created 100 different mock maps by injecting sources sampled from the model down to 1~mJy (as done in previous work, e.g. \citealt{Nowotka2022}) at random positions of the corresponding jackknife map. From these realizations, we obtained average number counts and used them to adjust the raw number counts by multiplying with the ratio of raw counts to mean recovered counts. The simulation then repeats from the Schechter function fitting now using the adjusted points and only stops once the mean recovered counts are within 1$\sigma$ of the raw counts in three consecutive loops (green versus black data points in Fig.~\ref{fig:diffnums}). For some fields (e.g. ID 3), the highest flux bin had to be excluded from the fit to achieve convergence due to small number statistics.
The parameters of the converged Schechter functions are summarized in Table~\ref{tab:Schechter}.

\begin{table*}
\caption{Parameters of the Schechter functions reproducing the observed source counts, galaxy overdensities calculated from the models, and angle direction of the overdensities.}
\label{tab:Schechter}
\centering 
\renewcommand{\arraystretch}{1.3}
\begin{tabular}{c c c c c c c c c c}
\hline\hline
ID & S/N\tablefootmark{a} & $N_0$ [deg$^{-1}$] & $\gamma$ & $\delta_{\rm scaled}$ & $\delta_{\rm cumul}$ & SFRD$_{\rm int}$\tablefootmark{b} & SFRD$_{\rm cen}$\tablefootmark{b} & $\alpha_{\rm RMS}$\tablefootmark{c} & $\alpha_{\rm inertia}$\tablefootmark{c} \\    
\hline
1 & 2.8 & 23227 $\pm$ 13462 & 1.00 $\pm$ 0.59 & $5.1 \pm 0.9$ & $4.4^{+2.5}_{-1.8}$ & 1000 $\pm$ 600 & 3000 $\pm$ 1900 & $14^{+36}_{-83}$ & $24^{+33}_{-61}$ \\
2 & - & - & - & - & - & - & - & - & - \\
3 & 2.7 & 27998 $\pm$ 6866 & 2.42 $\pm$ 0.32 & $1.8 \pm 0.2$ & $2.8^{+0.7}_{-0.5}$ & 2200 $\pm$ 200 & 27300 $\pm$ 17100 & $53^{+23}_{-28}$ & $55^{+23}_{-28}$ \\
4 & 3.3 & 14447 $\pm$ 13271 & 1.00 $\pm$ 0.82 & $3.4 \pm 0.9$ & $2.8^{+3.0}_{-1.8}$ & 800 $\pm$ 800 & 2800 $\pm$ 1800 & $50^{+22}_{-18}$ & $48^{+52}_{-24}$ \\
5 & 2.5 & 37265 $\pm$ 9191 & 1.88 $\pm$ 0.33 & $3.9 \pm 0.4$ & $4.6^{+1.1}_{-0.9}$ & 2400 $\pm$ 300 & 18200 $\pm$ 11400 & $37^{+37}_{-78}$ & $53^{+19}_{-24}$ \\
6 & 2.6 & 27432 $\pm$ 5876 & 1.67 $\pm$ 0.33 & $3.5 \pm 0.4$ & $3.7^{+0.7}_{-0.6}$ & 1700 $\pm$ 200 & 2500 $\pm$ 1500 & $78^{+67}_{-46}$ & $37^{+28}_{-25}$ \\
ELAN0101 & 3.1 & 16615 $\pm$ 7643 & 1.30 $\pm$ 0.62 & $2.7 \pm 0.5$ & $2.5^{+1.1}_{-0.8}$ & 700 $\pm$ 300 & 1000 $\pm$ 600 & $110^{+14}_{-54}$ & $136^{+9}_{-145}$ \\
J0119 & 3.1 & 21653 $\pm$ 15172 & 1.00 $\pm$ 0.67 & $5.0 \pm 1.0$ & $4.3^{+3.3}_{-2.1}$ & 1200 $\pm$ 800 & 200 $\pm$ 100 & $52^{+10}_{-61}$ & $26^{+13}_{-44}$ \\
J1622 & 2.8 & 43693 $\pm$ 15572 & 2.14 $\pm$ 0.45 & $3.6 \pm 0.4$ & $4.5^{+1.5}_{-1.2}$ & 2900 $\pm$ 600 & 50800 $\pm$ 31900 & $63^{+59}_{-15}$ & $52^{+23}_{-19}$ \\

\hline
\end{tabular}
\tablefoot{
\tablefoottext{a}{SNR threshold used for the extraction of differential number counts}
\tablefoottext{b}{in M$_{\odot}$~yr$^{-1}$~Mpc$^{-3}$}
\tablefoottext{c}{in degree, measured anti-clockwise with respect to the x-axis}
}
\end{table*}

We further verified that the source model found through Monte Carlo simulations properly reproduces the observed data including all observational biases by creating 500 mock maps per field and comparing the recovered source count with the observed counts.
Finally, we created 500 mock maps of the field by injecting sources according to the field model found by \cite{Geach2017} (fiducial model)  with $N_0 = 7180 \pm 1220$~deg$^{-2}$, $S_0 = 2.5 \pm 0.4$~mJy, and $\gamma = 1.5 \pm 0.4$. In that work, the blank field was derived from almost 3000 submillimeter source detections in 850~$\mu$m SCUBA-2 maps covering an area of roughly 5~deg$^2$. This test instructs us on whether the blank field source model is consistent with our observed differential number counts or not.
The differential number counts and confidence intervals obtained from the 500 realizations as well as the underlying source models are displayed in Fig.~\ref{fig:diffnums}.
Additionally, we used the 500 simulated maps of the underlying source model per field to quantify the flux boosting and positional uncertainty $\Delta(\alpha,\delta)$ of the source extraction at 850~$\mu$m by comparing the position and flux of the injected and recovered sources. Since we do not recover an underlying source model at 450~$\mu$m, we instead injected sources according to the model found by \cite{Casey2013} to obtain the boosting factor and positional uncertainty.

\section{Results}
\label{sec:results}

\subsection{Galaxy overdensity factors}
\label{subsec:results:overdens}

From the reconstructed underlying source models (Fig.~\ref{fig:diffnums}), the overabundance of SMGs in the field of quasar pairs in comparison to the fiducial model \citep{Geach2017} is evident. To quantify the degree of this effect, we calculated galaxy overdensity factors based on the model normalization ($\delta_{\rm scaled}$) and the cumulative differential number counts ($\delta_{\rm cumul}$) as previously done in \cite{Nowotka2022}.

Specifically, to obtain $\delta_{\rm scaled}$, we scaled the observed differential number counts by the ratio between the underlying model and the mean recovered number counts to correct for observational biases and fitted the field model from \cite{Geach2017} to the resulting points, using the symmetrized error bars as weights and leaving the normalization factor $N_0$ free to vary. The ratio between normalization factors of the quasar pair fields and the fiducial model is then $\delta_{\rm scaled}$. The obtained values have a weighted average of $2.7 \pm  0.1$ and are compiled in Table~\ref{tab:Schechter}.
This method is particularly successful if the slopes of the Schechter functions are similar to the fiducial model (e.g. ID 6 and ID ELAN0101), but does not capture the full picture for fields with values of $\gamma$ very different from $\gamma = 1.5$.
Thus, we additionally calculated the cumulative overdensities $\delta_{\rm cumul}$ by integrating the source models between the minimum and maximum flux of detected sources and determining the ratio to the fiducial model integrated in the same range. To obtain a robust value and the typical error range, we randomly vary all free parameters of the two models within their 1$\sigma$ error 50,000 times. The values for $\delta_{\rm cumul}$ and their errors are then obtained from the median, 16th, and 84th percentile of the resulting distribution and also listed in Table~\ref{tab:Schechter}. By symmetrizing the error, we obtained the weighted average of $3.4 \pm 0.3$ for the overdensity factor determined in this way.
The implication of the calculated overdensity factors are discussed in Sec.~\ref{sec:dis:protocluster}.

\subsection{Star formation rates and molecular gas masses}
\label{subsec:results:sfr}

\begin{figure}
    \centering
    \includegraphics[width=0.99\hsize]{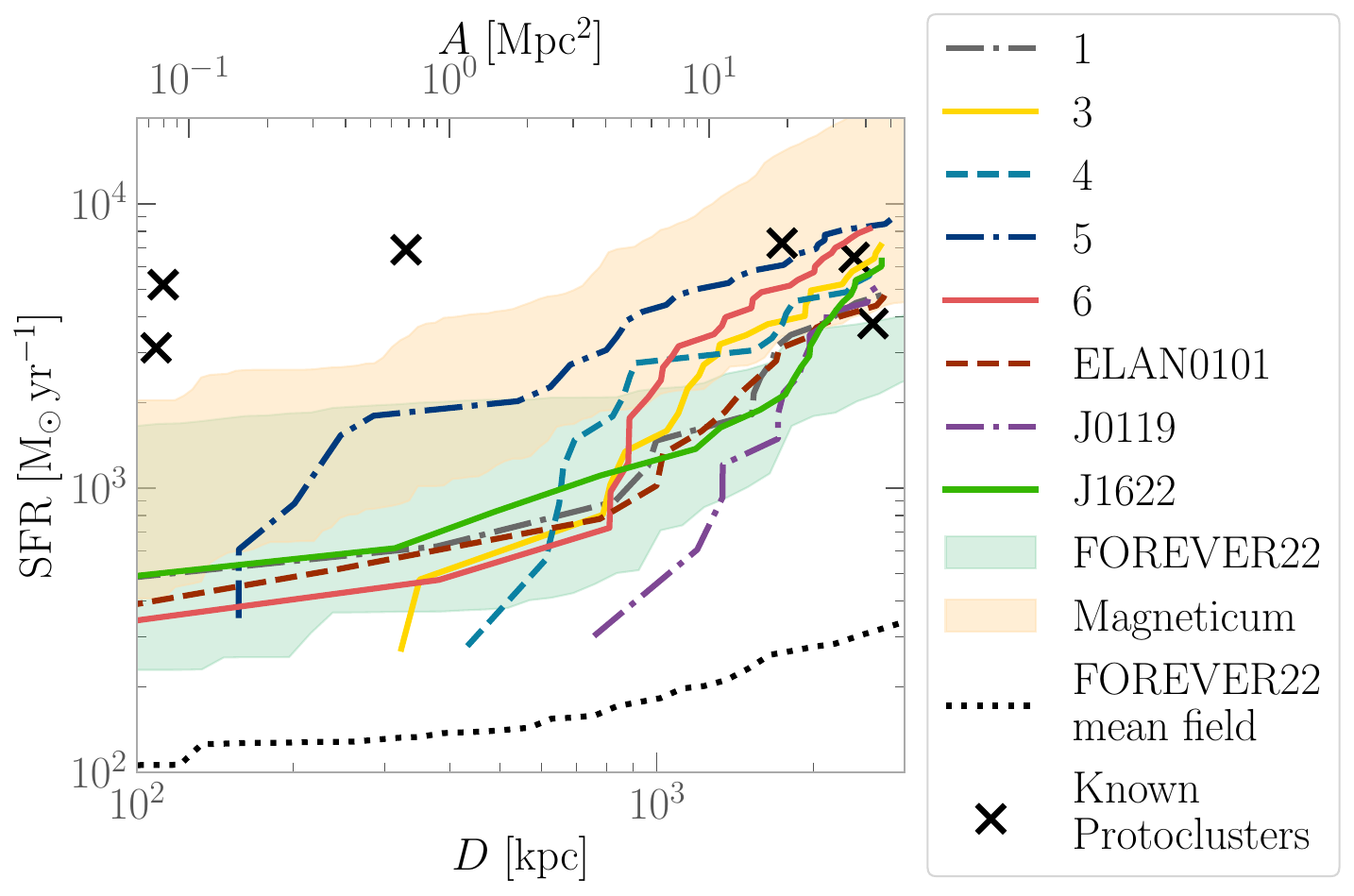}
    \caption{Cumulative SFR of the quasar pair fields compared to the ten protocluster zoom-in regions simulated in FOREVER22 at $z = 3$ \citep{Yajima2022}, the 32 most massive halos from Magneticum at $z=2.79$ \citep{Remus2023, Dolag2025}, and known protoclusters from the literature (\citealt{Wang2025} and references therein). The dotted line shows the cumulative SFR of the mean blank field in FOREVER22 \citep{Yajima2022}.}
    \label{fig:cumsfr}
\end{figure}

\begin{figure}
    \centering
    \includegraphics[width=0.99\hsize]{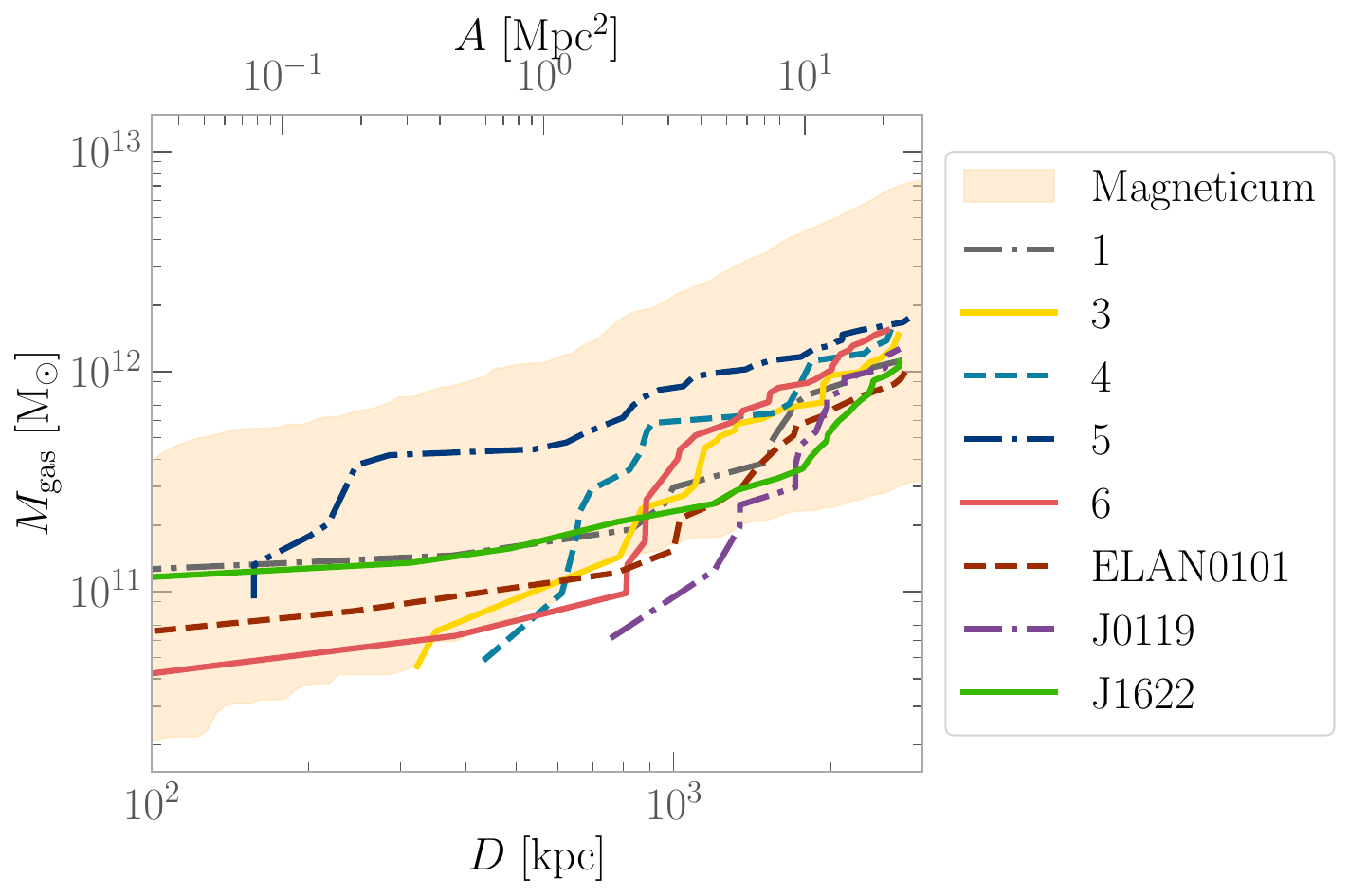}
    \caption{Cumulative molecular gas mass $M_{\rm gas}$ of the quasar pair fields compared to the molecular gas mass in the 32 most massive halos in Magneticum at $z=2.79$ \citep{Remus2023, Dolag2025}, estimated from the star forming gas.}
    \label{fig:mgas}
\end{figure}

Because the dust temperature and therefore the emission at submillimeter wavelengths is driven by the obscured star formation, we can calculate the SFR and dust mass $M_d$ for each source from its 850~$\mu$m flux. We use the relations found by \cite{Dudzeviciut2020} based on ALMA data and UV-to-radio SED fitting of 707 SMGs,
selected from the SCUBA-2 Cosmology Legacy Survey \citep{Geach2017},
    ${\rm log(SFR/(M_{\odot}\, yr^{-1}))} = (0.42 \pm 0.06) \times {\rm log}(S_{870}/{\rm mJy}) + (2.19 \pm 0.03)$ 
    and ${\rm log(M_d/(M_{\odot}))} = (1.20 \pm 0.03) \times {\rm log}(S_{870}/{\rm mJy}) + (8.16 \pm 0.02)$.
We calculated the cumulative SFRs and molecular gas masses $M_{\rm gas}$, assuming a gas-to-dust ratio of 63 \citep{Birkin2021}, of all SMG candidates and quasar detections with respect to the coordinate in the middle between the quasar pair for each field and compared the SFRs to literature protoclusters (\citealt{Wang2025} and references therein), the cumulative SFRs of the ten simulated protocluster regions in FOREVER22 \citep{Yajima2022} at $z=3$ and the 32 most massive halos at the respective redshift from the Magneticum Pathfinder simulation \citep{Remus2023, Dolag2025}. The SFRs from simulations include all types of star forming galaxies and are not limited to massive, dusty galaxies equivalent to SMGs or flux-limited in agreement with the observations.
To compare to the gas mass in the 32 Magneticum halos, we assumed a molecular-to-atomic gas ratio between 10~\%, corresponding to the lowest estimates for small galaxies \citep{Saintonge2011, Catinella2018}, and 70~\%, found in highly star forming, massive galaxies \citep{Saintonge2016}.

FOREVER22 is a set of zoom-in hydrodynamical simulations of protocluster regions selected at $z=2$ from the parent N-body box of 714~cMpc per side and each simulated in a box size of (28.6~cMpc)$^3$ until $z_{\rm end}=2$. At $z=0$, the selected halos span masses between $5.2 \times 10^{14}$~M$_{\odot}\, h^{-1}$ and $1.4 \times 10^{15}$~M$_{\odot}\, h^{-1}$. The initial gas particle mass is set to $2.9 \times 10^6$~M$_{\odot}\, h^{-1}$ and the gravitational softening length is 2~ckpc~$h^{-1}$. 
Magneticum is a set of hydrodynamic simulations of cosmological boxes, including Box2b of 640~Mpc~$h^{-1}$ per side 
from which the displayed halos are selected. It is simulated to $z=0.2$ with a gas particle mass of $1.4 \times 10^{8}$~M$_{\odot}\, h^{-1}$ and a gravitational softening length of 3.75~kpc~$h^{-1}$. The virialized halo masses of the displayed protocluster regions at $z=2.79$ span masses log($M$/M$_\odot h^{-1}) = 13.6 - 14.1$ (log($M$/M$_\odot h^{-1}) = 13.4 - 14.3$ for all four redshifts). Although not all the selected regions will collapse into galaxy clusters by $z=0$, the likelihood increases with decreasing redshift and almost all halos selected at $z=2$ end as one in the last simulation snapshot \citep{Remus2023}.

The resulting cumulative SFRs are shown in Fig.~\ref{fig:cumsfr}, and the cumulative $M_{\rm gas}$ are displayed in Fig.~\ref{fig:mgas}. Because the calculations rely on source detections, our observational values are heavily influenced by the data quality as the completeness of the source detection decreases under worse observing conditions and shorter exposure time,  and we thus attempted to correct for these effects when converting SFRs to SFR densities in two different ways.
Firstly, we calculated SFRD$_{\rm int}$ by integrating the obtained underlying source model (Table~\ref{tab:Schechter}) multiplied with the SFR relation in the range $S_{850} = (1-20)$~mJy and weighting the result by the volume of a sphere with a projected area of $A_{\rm eff}$ (Table~\ref{tab:obslog}). The 16th, 50th and 84th percentile range give SFRD$_{\rm int} = 1400_{-600}^{+1000}$, and the weighted average is SFRD$_{\rm int} = 1700 \pm 100$. Given the completeness curve and assuming a uniform source model throughout the structure, we would expect to detect only up to 12~\% of the SFR derived in this way.

\begin{figure}
    \centering
    \includegraphics[width=0.99\hsize]{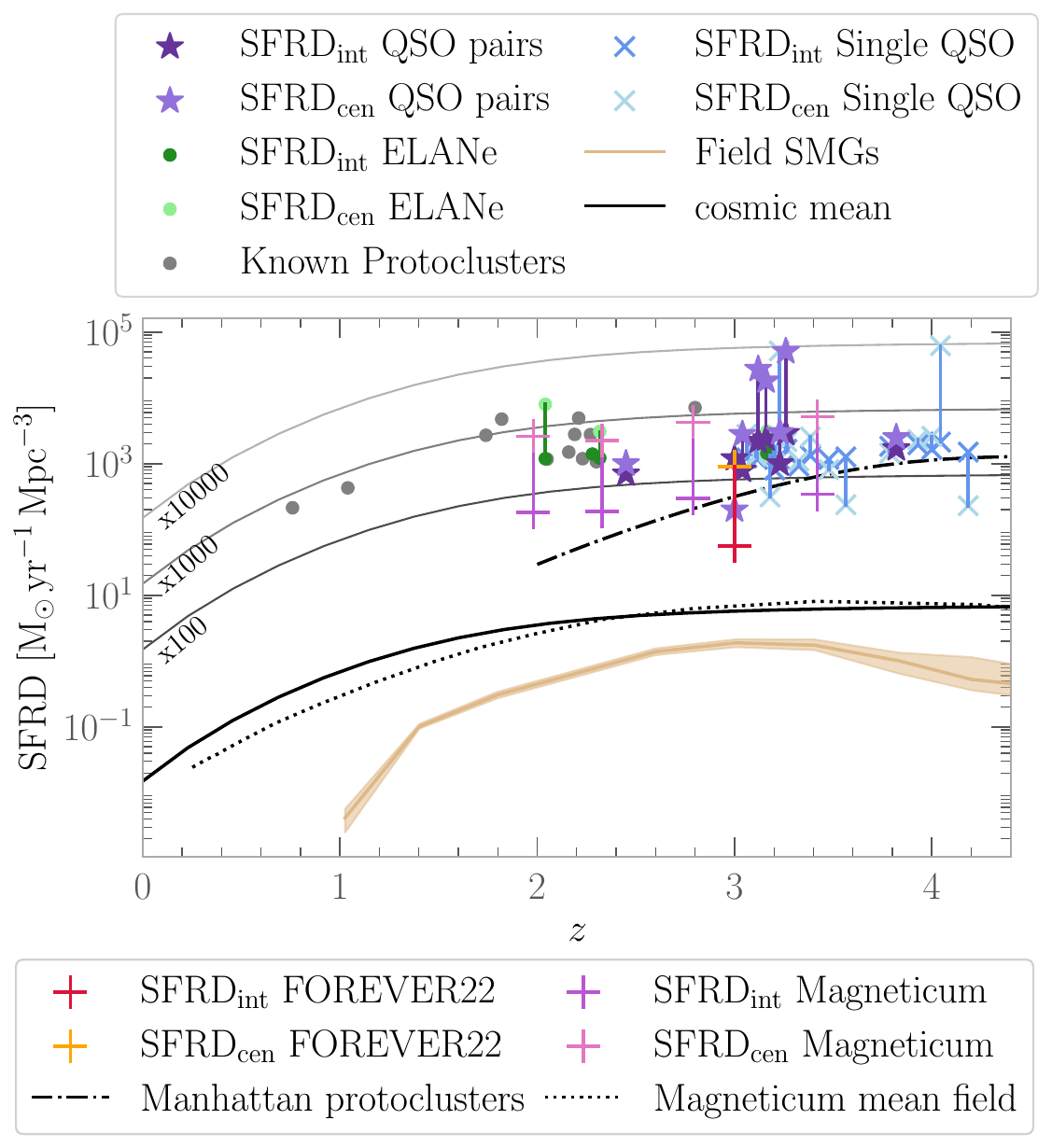}
    \caption{The SFRD in the central 1~Mpc (SFRD$_{\rm cen}$) and from the source model of the full field (SFRD$_{\rm int}$) for quasar pairs, ELANe \citep{Nowotka2022, FAB2023} and single quasars \citep{FAB2023}, compared to values reported for known protoclusters \citep{Clements2014, Dannerbauer2014, Kato2016} and simulation predictions from FOREVER22 (PCR0; \citealt{Yajima2022}) and Magneticum Box2b (the most massive halo; \citealt{Remus2023}), both integrated within the central 1~Mpc (SFRD$_{\rm cen}$) and 3~Mpc (SFRD$_{\rm int}$). The solid black line shows the mean cosmic SFRD \citep{Madau2014}, the solid grey lines are the mean cosmic SFRD scaled up by a factor of 100, 1000, and 10000 respectively, the beige line shows the SFRD for blank field SMGs \citep{Dudzeviciut2020}, the dash-dotted line shows the simulation prediction for SFRD within protoclusters by \cite{Rennehan2024}, and the dotted line is the SFRD in the Magneticum mean field \citep{Dolag2025}.}
    \label{fig:sfrd}
\end{figure}

We then calculated SFRD$_{\rm cen}$ by only summing up source detections, including SMG candidates and quasar detections, within the central 1~Mpc and weighting them by the completeness within this smaller aperture and the volume of a sphere with 1~Mpc radius. The median value is SFRD$_{\rm cen} = 2900_{-1800}^{+23200}$, while the weighted average is dominated by fields with few central detections and therefore low total error with SFRD$_{\rm cen}= 204 \pm 97$.

A flux-SFR conversion relation based on ALMA data can lead to further uncertainties if applied to SCUBA-2 data due to the difference in beam size. We thus calculated the SFRs using the relation reported in \cite{Cowie2017} based on photometrically selected SMGs and find higher values for every field with an average increase of 45~\% in SFRD$_{\rm int}$ and 75~\% in SFRD$_{\rm cen}$. However, we opt to report the more conservative values derived from redshift-confirmed SMGs and compile the results in Table~\ref{tab:Schechter}.

For comparison, we also calculated SFRD$_{\rm int}$ and SFRD$_{\rm cen}$ for the source models of single quasar fields (median values with 16th and 84th percentile ranges of SFRD$_{\rm int} = 1300_{-200}^{+500}$ and SFRD$_{\rm cen} = 2400_{-1500}^{+1100}$, weighted average SFRD$_{\rm int} = 1700 \pm 100$ and SFRD$_{\rm cen}= 400 \pm 100$) and ELAN fields reported in \cite{FAB2023}, and calculate SFRD$_{\rm int}$ and SFRD$_{\rm cen}$ for PCR0 in FOREVER22 and the most massive halo of the Magneticum Box2b by summing up sources within 3~Mpc (i.e. the typical radius obtained from $A_{\rm eff}$ in the quasar pair fields) and 1~Mpc respectively. The results are shown in Fig.~\ref{fig:sfrd}.
We also compare to the redshift evolution of the cosmic mean SFRD (MD14; \citealt{Madau2014}), the mean field SFRD in Box2b of Magneticum \citep{Dolag2025}, and predictions of the mean SFRD in protoclusters from the Manhatten suite of 100 zoom-in simulations of protocluster regions with $M > 10^{14}$~M$_{\odot}$ selected at $z=2$ as the most massive halos from an N-body box with a side length of 1.5~cGpc \citep{Rennehan2024}. The zoom-in areas include $9\times$R$_{\rm vir}$ at a gas particle mass resolution of roughly $3.5 \times 10^7$~M$_{\odot}$ and an adaptive gravitational softening length, but a minimum value of 0.73~ckpc.
Overall, Figures~\ref{fig:cumsfr} to \ref{fig:sfrd} show that although our candidate SMGs represent only a fraction of the total SFR and cold gas
mass in simulated protocluster structures because of the exclusion of less obscured star forming galaxies, our measurements are in agreement with model predictions when considering potential incompleteness of the observations.
We further discuss the derived values for SFR and SFRDs from our observations and their connection to simulation values in Sec.~\ref{sec:dis:protocluster}.

\subsection{Direction of overdensity}
\label{subsec:results:angle}

\begin{figure*}[h]
    \centering
    \includegraphics[width=0.87\hsize]{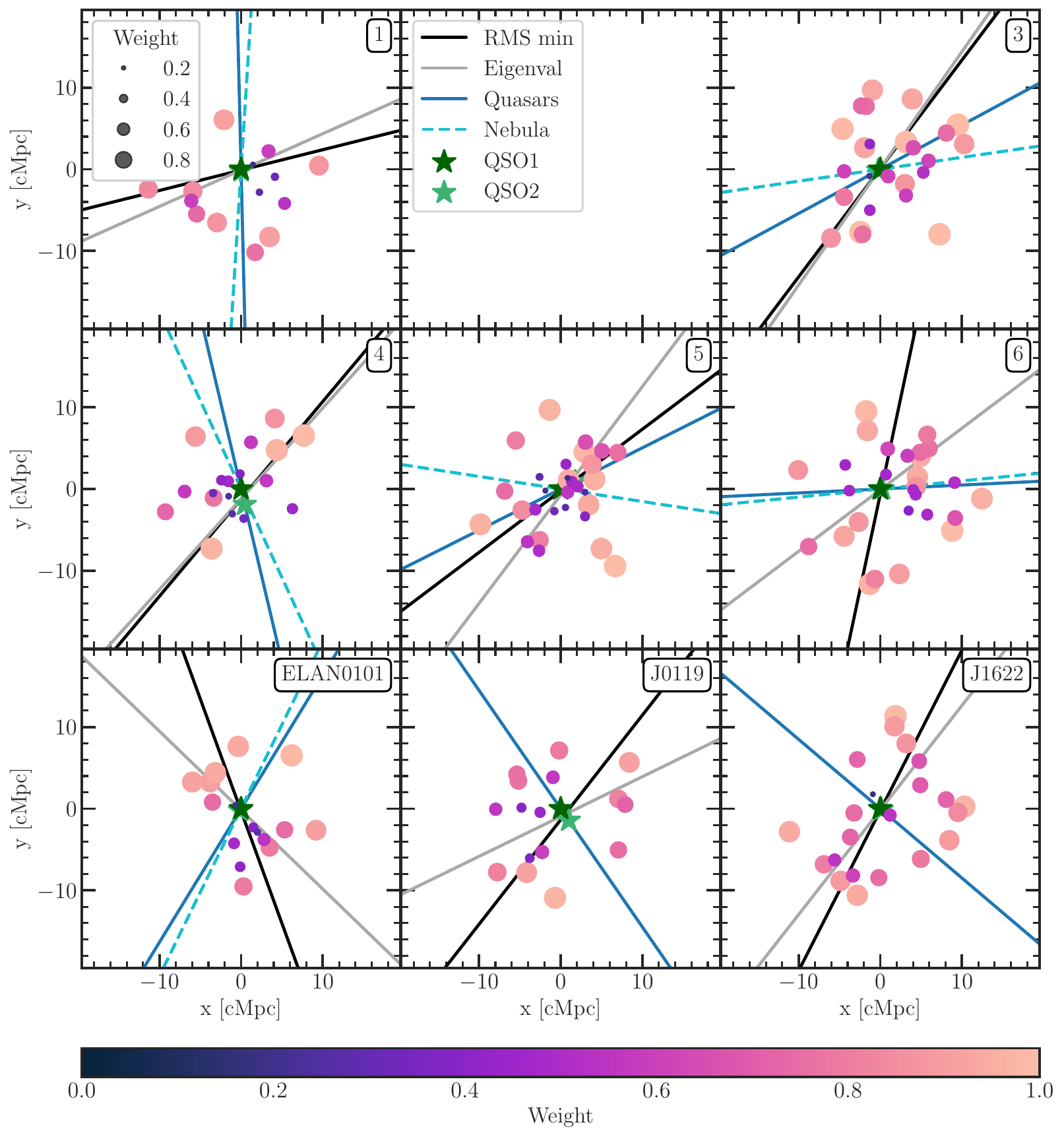}
    \caption{Direction of the overdensity, calculated from the eigenvalue and the minimized rms, as well as the line connecting the quasar pair and the semi-major axis of the nebula, if available. All catalog sources of the fields are plotted, size and color of the points indicate the weight according to catalog completeness at the flux of the SMG (i.e., within one field, larger and brighter points correspond to higher fluxes).}
    \label{fig:direction}
\end{figure*}

We determined the preferred direction of the galaxy overdensity in each field using two different methods following \cite{FAB2023}. If massive galaxies such as SMGs are tracers of filaments connecting to the quasar in the center, this direction would be equivalent to the projection of a cosmic web filament.
Using the coordinate in the middle between the quasars as a reference point, we first calculated the angle $\alpha_{\rm RMS}$ (measured from East to North) that minimizes the rms distance of detected sources weighted by the completeness. For robust results, we performed the calculation in 50,000 randomly drawn subsamples encompassing 90~\% of the sources with replacement, smoothed the obtained angle distribution using a Gaussian kernel density estimate and defined the peak as $\alpha_{\rm RMS}$. We obtained errors from the 16th and 84th percentile.
Second, we determined the angle $\alpha_{\rm inertia}$ by computing the 2D inertia tensor from the source positions weighted by completeness and obtaining the eigenvector that minimizes the eigenvalue \citep{Zeballos2018}. Equivalently to the other method, we used the peak value and 16th and 84th percentile of the angle distribution from 50,000 90~\% subsamples to get $\alpha_{\rm inertia}$ and the corresponding errors. In all cases, $\alpha_{\rm RMS}$ and $\alpha_{\rm inertia}$ are consistent with each other within error bars. The values are compiled in Table~\ref{tab:Schechter} and the preferred SMG directions and source weights are displayed in Fig.~\ref{fig:direction}, together with the line connecting the two quasars and the direction of the semi-major axis of the Ly$\alpha$ nebula, if available from \cite{Herwig2024}. For ELAN0101, we obtained the nebula semi-major axis analogously to the calculation in \cite{Herwig2024} from Figure~1 (upper right panel) in \cite{Cai2018}.
 We note that choosing one of the quasars as reference point instead does not significantly impact the obtained angles. Specifically, $\alpha_{\rm RMS}$ deviates by less than 6$^{\circ}$, while $\alpha_{\rm inertia}$ can differ by up to 12$^{\circ}$.
In Appendix~\ref{app:testangle}, we show the success in recovering the angle direction of simulated SMG distributions within filaments of different widths.
The calculated angles and related findings are discussed in Sec.~\ref{sec:dis:filaments}.

\section{Discussion}
\label{sec:discussion}

\subsection{Do quasar pairs trace protocluster environments?}
\label{sec:dis:protocluster}

Contrary to previous studies of single AGN fields \citep{Rigby2014, Jones2017, FAB2023}, we find an increased density of SMGs in all objects of our sample of quasar pair fields and observe a cumulative SFR and SFRD roughly in agreement with known protoclusters from the literature and simulations and, compared to the field, increased by a factor of 100 to 10,0000. If the majority of detected sources are truly associated with the quasar pairs, these findings strongly support the hypothesis that quasar pairs indeed trace protocluster environments. While we attempted to exclude interlopers using the 450~$\mu$m band, fainter interloping galaxies undetected in this band can still contaminate the sample and follow-up observations are necessary for definitive redshift confirmation of the SMG candidates. On the other hand, SCUBA-2 observations using the daisy scan pattern lead to stark differences in the sensitivity throughout the field as the noise quickly increases towards the edges of the image. Thus, employing this observational strategy, we would likely miss important parts of the galaxy population in our SMG candidate catalogs if the quasar pair is offset from the center of a forming galaxy protocluster, and imaging with uniform sensitivity throughout the likely protocluster extent of roughly 20~cMpc would alleviate this problem.

An offset between the reference point of our observations and the true protocluster center could potentially also explain why most simulated protoclusters show higher cumulative SFRs in the center compared to the quasar pair fields observed here (Fig.~\ref{fig:cumsfr}), though quasar pairs still significantly exceed the mean field expectation from FOREVER22. Alternatively, the central SFR might not be dust obscured in all quasar pair fields and thus would not appear at submillimeter wavelengths, while the SFRs obtained for simulated protoclusters include all types of star forming galaxies and are therefore expected to be shifted towards higher values. On the other hand, most cosmological simulations are known to not reproduce the extremely bursty SFR observed in SMGs well \citep{Bassini2020, Lim2021, Remus2023}, leading to a potential suppression of the cumulative SFR.
At larger radii, the cumulative SFR of the quasar pair fields exceeds the predictions from the FOREVER22 simulations, but most data points stay below the cumulative SFRs in Magneticum protoclusters, with the exception of quasar pair~5, the field hosting a candidate quadruple AGN in the center \citep{Herwig2025}, which fits the radial profile of the Magneticum predictions well. On the other hand, the cumulative molecular gas masses $M_{\rm gas}$ of the observed quasar pairs are in good agreement with the Magneticum predictions (Fig.~\ref{fig:mgas}), though the uncertainty in molecular-to-ionized gas ratio and gas-to dust-ratio are large. The central cumulative SFR around quasar pairs is also lower than in other known protoclusters, although all but one are indeed dominated by central SFR within 1~Mpc from the quasar pair (Fig.~\ref{fig:sfrd}). This is in agreement with ELANe, while single quasar fields are instead dominated by extended SFR in more than a third of the fields reported in \cite{FAB2023}.
Large cool gas reservoirs are expected to be a necessary requirement to drive both ELANe and AGN activity in multiple quasars, and could feasibly feed an abundance of SFR in the central region of a protocluster simultaneously, explaining the excess of SFRD$_{\rm cen}$ compared to SFRD$_{\rm int}$.
The SFRD for spectroscopically confirmed blank field SMGs \citep{Dudzeviciut2020} compared to the cosmic mean SFRD \citep{Madau2014} shown in Fig.~\ref{fig:sfrd} further highlights the contribution from non-SMGs to the SFRD and therefore likely the SFR in protocluster regions and the necessity for multi-wavelength observations of the quasar pair fields presented in this work for a comprehensive view of the potential protocluster regions and a better comparison to simulations.
The central SFRD in the Magenticum and FOREVER22 simulations are of the same order of magnitude as observed protoclusters and protocluster candidates, while the SFRD integrated over larger areas tend towards lower values, especially for the FOREVER22 zoom-ins and as already seen in Fig.~\ref{fig:cumsfr}. Additionally, Magneticum reproduces the mean field SFRD well. Predictions for the SFRD within all 100 protoclusters of the Manhattan suite of simulations fit observational data points above $z \sim 3$, but show a steeper decline towards low redshifts. This could be a consequence of their selection based on mass at $z = 2$, which might require a starburst peak at higher redshifts to build up sufficient mass.

The presented average SMG overdensity of $\delta_{\rm cumul} = 3.4 \pm 0.3$ in quasar pair fields significantly exceeds the average value of $\delta_{\rm cumul} = 2.5 \pm 0.2$ found around single quasars with a similar approach as employed here \citep{FAB2023}, but is consistent with previously reported values for fields with multiple known quasars, $\delta_{\rm cumul} = 3.6 \pm 0.6$ \citep{Nowotka2022} and $\delta_{\rm cumul} = 3.4 \pm 0.4$ \citep{FAB2023}. While half of our sample has higher overdensity factors than the peak value of $\delta_{\rm cumul} = 3.9$ reported in both works, this is possibly a statistical effect.
The two studies targeted the environment of quasars associated with ELANe and concluded that these extreme Ly$\alpha$ nebulae are suitable tracers of galaxy overdensities.
In contrast to that, we find the lowest overdensity factor in the only field of our sample with a known ELAN (ELAN0101) with $\delta_{\rm cumul} = 2.6 \pm 0.9$. Instead, our findings suggest that the galaxy overdensities are causally connected to AGN overdensities, which are typically needed to power exceptionally far extended Ly$\alpha$ emission and are thus present in all ELANe fields. As such, we find the highest overdensity factor of $\delta_{\rm cumul} = 4.6 \pm 1.7$ around the quadruple AGN candidate (ID5), further highlighting this system as the most interesting field for follow-up studies and the most likely protocluster region of our sample.
In most fields (six out of eight), at least one quasar of the pair is detected above 3$\sigma$ at 850~$\mu$m, but the quadruple AGN candidate is associated with the brightest emission from any quasars in both bands. It also stands out as the only system of the sample with a radio-loud quasar detected in VLA FIRST \citep{Becker94}, though the radio emission location is consistent with the position of the brighter quasar and the two AGN candidates accompanying it \citep{Herwig2025}.

\subsection{Comparison of different methods to infer filament direction}
\label{sec:dis:filaments}

\begin{figure}
    \centering
    \includegraphics[width=\hsize]{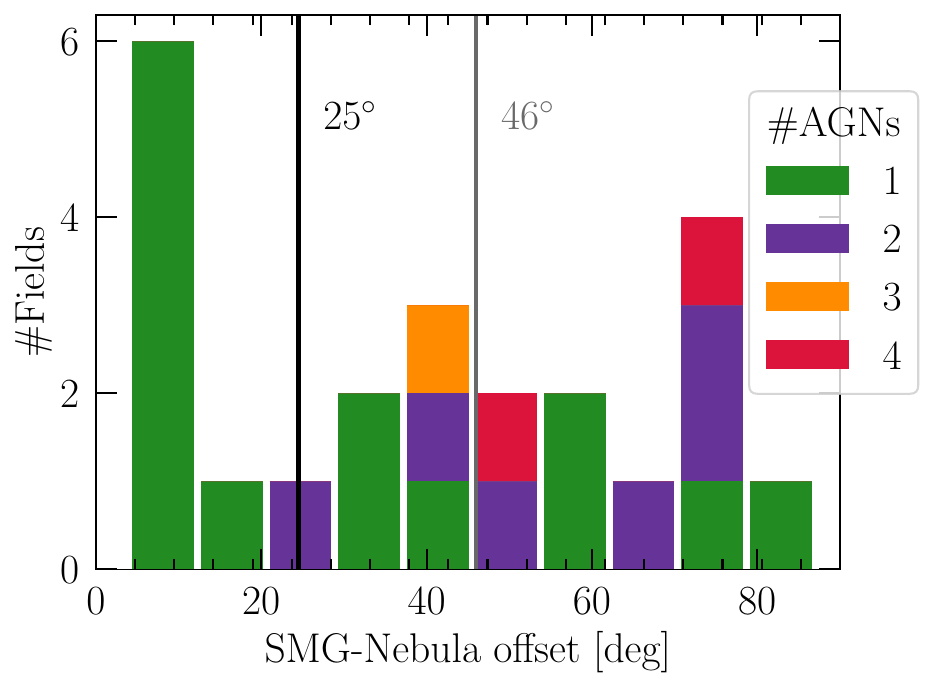}
    \caption{Histogram of the angle difference between semi-major axis of the nebula and direction of the Megaparsec-scale SMG distribution $\alpha_{\rm RMS}$, color-coded by the number of quasars in the center of the field. The angles are compiled from \cite{FAB2023}, \cite{Herwig2024} and Table~\ref{tab:Schechter}, and restricted to the range [$0^{\circ}$,$90^{\circ}$). The median angle offset of $25^{\circ}$ ($46^{\circ}$) for single (multiple) quasar fields is marked by the vertical black (grey) line.}
    \label{fig:anglehist}
\end{figure}

We attempted to infer the direction of cosmic web filaments in three different ways for most fields: the preferred alignment direction of detected SMG candidates ($\alpha_{\rm RMS}$, $\alpha_{\rm inertia}$), the line connecting the two quasars as they are likely connected by a filament due to their proximity and large halo mass, and the semi-major axis of the Ly$\alpha$ nebulae, if available from \cite{Herwig2024}, as the extended emission might trace cool gas accreted onto the circumgalactic medium from filaments connecting to the halo \citep{Costa22}. While \cite{Herwig2024} found alignment of the latter two values for all Ly$\alpha$ nebulae extending for more than 70~kpc from the brighter quasar (ID1, ID4, ID6), these angles do not coincide with the preferred direction of SMGs. In fact, in most cases and irrespective of projected quasar pair separation, pair direction and Ly$\alpha$ emission direction are almost perpendicular to the SMG direction (ID1, ID4, ID6 for $\alpha_{\rm RMS}$, ELAN0101, J0119, J1622). Only two fields (ID3 and ID5) show signs of alignment of the Mpc-scale galaxy distribution and the kiloparsec-scale quasar pairs or their extended Ly$\alpha$ emission. This finding is in agreement with \cite{FAB2023}, who found alignment between SMGs and Ly$\alpha$ nebulae for single quasar fields, but average offset angles of $\alpha_{\rm RMS} = 41^{+16}_{-17}$ and $\alpha_{\rm inertia} = 68^{+39}_{-34}$ for ELANe fields (i.e. fields with multiple quasars). To further highlight this point, we show the distribution of angle offsets between SMG distribution and Ly$\alpha$ nebula, restricted to angles between $0^{\circ}$ and $90^{\circ}$, for different known quasar number per field in Fig.~\ref{fig:anglehist}. While the median values with 16th and 84th percentile error ($25^{+36}_{-15}$ degrees and $46^{+26}_{-5}$ degrees for single quasar fields and multiple quasar fields respectively) are consistent within errors of each other, the most dominant bin for single quasar fields spans from 0 degree to 10 degree, while most multiple-AGN fields can be found at an offset angle between 70 and 80 degrees.

A possible cause for the misalignment could be the increased importance of galaxy interactions for the Ly$\alpha$ nebula morphology in the dense regions of quasar pairs. In this picture, the angles measured from quasars and extended Ly$\alpha$ emission might trace small-scale dense gas filaments shaped by a potential ongoing galaxy merger, while SMGs trace large-scale filaments which would only sometimes coincide in projection due to a random occurrence. Further, the connectivity of halos to the cosmic web increases with increasing halo mass \citep{Galarraga-Espinosa2024}. In other words, more cosmic web filaments are expected to be linked to a halo the higher its mass. If two quasars are physically associated, their combined halo mass would increase the likelihood of a complicated structure of gas filaments in their direct environment, and their directions might not be simply described by a straight line determined from the surrounding galaxy population. Further complicating the geometry of filaments within protocluster regions, the most massive SMGs have similarly high halo masses as quasars and therefore likely reside in nodes of the cosmic web themselves. However, the majority of the population consists of less massive galaxies that should preferentially lie along the spines of filaments.

\section{Summary and conclusions}
\label{sec:summary}

Using JCMT/SCUBA-2 850~$\mu$m imaging covering an area of 106~arcmin$^2$ on average (or 23.1~Mpc$^2$ at the median $z = 3.13$), we searched for submillimeter sources around nine quasar pairs at $z = 2.45 - 3.83$. Our main findings are as follows:

\begin{itemize}
    \item We identified 170 SMG candidates at 4$\sigma$ significance in the 850~$\mu$m band or at 3$\sigma$ significance in both the 850~$\mu$m and 450~$\mu$m band, 26 interloper or non-SMG candidates, and eight detections >3$\sigma$ for quasars.
    \item All objects targeted in this work are surrounded by SMG overdensities when compared to blank fields, with an average overdensity factor of $\delta_{\rm cumul} = 3.4 \pm 0.3$, confirming that within our sample, physically associated quasar pairs are reliable tracers of galaxy overdensities. $\delta_{\rm cumul}$ is similar to ELAN fields, and significantly higher than around single quasars.
    \item The cumulative SFR and SFRDs are consistent with theoretical predictions for protoclusters, known protoclusters and single quasar fields, although SFR around quasar pairs tends to be more centrally peaked when compared to single quasars.
    \item We do not find alignment between the direction of the SMG overdensity and the quasar pair direction or the semi-major axis of extended Ly$\alpha$ emission associated with the circumgalactic medium of the quasar pairs, indicating that SMGs might trace different, larger-scale filaments than the other metrics.
\end{itemize}

This work showcases the utility of quasar pairs in pinpointing dense megaparsec-scale environments. The overdensity factors and SFRDs identify the studied fields as likely sites of protocluster formation. However, spectroscopic follow-up studies of all SMG candidates proposed in this work are necessary to confirm the density and geometry of the structures surrounding quasar pairs. To get a complete view of the galaxy population around quasar pairs and test the filament direction inferred from the SMG distribution against other methods, multi-wavelength observations identifying smaller star forming galaxies such as Ly$\alpha$ emitters (e.g. \citealt{Garcia-Vergara2019}) or Lyman break galaxies \citep{Steidel2003} are necessary.

\begin{acknowledgements}
We thank Guinevere Kauffmann for providing useful comments to an earlier version of this work. This work made use of data taken under the Program IDs M22AP049 and M22BP025. These observations were obtained by the James Clerk Maxwell Telescope, operated by the East Asian Observatory on behalf of The National Astronomical Observatory of Japan; Academia Sinica Institute of Astronomy and Astrophysics; the Korea Astronomy and Space Science Institute; the National Astronomical Research Institute of Thailand; Center for Astronomical Mega-Science (as well as the National Key R\&D Program of China with No. 2017YFA0402700). Additional funding support is provided by the Science and Technology Facilities Council of the United Kingdom and participating universities and organizations in the United Kingdom and Canada. Additional funds for the construction of SCUBA-2 were provided by the Canada Foundation for Innovation. The authors wish to recognize and acknowledge the very significant cultural role and reverence that the summit of Maunakea has always had within the indigenous Hawaiian community.  We are most fortunate to have the opportunity to conduct observations from this mountain. Computations were performed on the FREYA cluster hosted by The Max Planck Computing and Data Facility (MPCDF) in Garching, Germany. We acknowledge the support provided by the MPCDF helpdesk. This work has made use of the Python libraries \texttt{matplotlib} \citep{Hunter2007}, \texttt{numpy} \citep{Walt2011}, \texttt{astropy} \citep{astropy:2013, astropy:2018, astropy:2022}, and scipy \citep{scipy2020}. C.-C.C. acknowledges support from the National Science and Technology Council of Taiwan (111-2112-M-001-045-MY3), as well as Academia Sinica through the Career Development Award (AS-CDA-112-M02). A.O.'s contribution to this work was made possible by funding from the Carl Zeiss Stiftung. H.Y. acknowledges support by MEXT/JSPS KAKENHI Grant Number 21H04489 and JST FOREST Program, Grant Number JP-MJFR202Z. The Magneticum Pathfinder simulations were performed at the Leibniz-Rechenzentrum with CPU time assigned to the Project \textit{pr83li}. This work was supported by the Deutsche Forschungsgemeinschaft (DFG, German Research Foundation) under Germany's Excellence Strategy EXC-2094\,--\,390783311. We are especially grateful for the support by M.~Petkova through the Computational Center for Particle and Astrophysics (C2PAP).
\end{acknowledgements}

\bibliographystyle{aa} 
\bibliography{lit.bib}

\begin{appendix}
\section{Quasars}
\label{app:quasars}

In this appendix we report the information on the targeted quasars and their obtained constraints at 850 and 450~$\mu$m (Table~\ref{tab:quasars}). 

\begin{center}
\begin{minipage}{\textwidth}
\captionof{table}{Quasar coordinates, magnitudes and submillimeter flux detections above 3$\sigma$ within $6\arcsec$ of the quasars. For non-detections, the forced flux at the quasar coordinates is reported and marked with an asterisk.}
\label{tab:quasars}
\centering 
\begin{tabular}{l c c c c c c c}
\hline\hline
ID & RA (J2000) & Dec (J2000) & $r$ [mag] & $f_{850}$ [mJy] & $f_{\rm 850, db}$ [mJy] & $f_{450}$ [mJy] & $f_{\rm 450, db}$ [mJy]\\    
\hline
1 & 00:01:40.6 & +07:09:54.0  & 18.4 & $9.1 \pm 1.2$ & $7.8 \pm 1.0$ & $31.2 \pm 12.8$ & $2.8 \pm 1.1$ \\
  & 00:01:40.6 & +07:09:47.8 & 24.6 & *3.8 $\pm$ 1.2 & *2.2 $\pm$ 0.7 & *12.5 $\pm$ 12.8 & *1.1 $\pm$ 1.2\\  \hline
2 & 00:18:07.4 & 16:12:57.6 & 21.1 & *2.2 $\pm$ 2.7 & - & *30.1 $\pm$ 15.5 & - \\
  & 00:18:08.1 & 16:12:50.8 & 21.9 & *-1.4 $\pm$ 2.7 & - & *-0.0 $\pm$ 15.4 & - \\ \hline
3 & 02:40:05.2 & -00:39:09.8 & 22.2 & *-0.8 $\pm$ 0.9 & *-0.3 $\pm$ 0.4 & *-3.7 $\pm$ 9.1 & *-0.6 $\pm$ 1.4\\
  & 02:40:05.7 & -00:39:14.1 & 22.7 & *1.3 $\pm$ 0.9 & *0.6 $\pm$ 0.4 & *-14.6 $\pm$ 9.1 & *-2.2 $\pm$ 1.4\\ \hline
4 & 02:44:42.6 & -00:23:20.4 & 20.1 & $4.8 \pm 1.3$ & $3.0 \pm 0.8$ & *14.4 $\pm$ 14.1 & *0.9 $\pm$ 0.8\\
  & 02:44:41.7 & -00:24:20.2 & 22.3 & *0.0 $\pm$ 1.3 & - & *11.7 $\pm$ 14.3 & *0.7 $\pm$ 0.9\\ \hline
5 & 10:12:54.7 & 03:35:48.8 & 21.9 & $13.1 \pm 0.9$ & $11.6 \pm 0.8$ & $46.7 \pm 7.1$ & $36.4 \pm 5.5$ \\
  & 10:12:51.1 & 03:36:16.6 & 21.7 & *0.0 $\pm$ 1.0 & - & *9.5 $\pm$ 7.5 & 2.0 $\pm$ 1.6 \\ \hline
6 & 10:21:17.0 & 11:12:27.6 & 20.6 & $3.4 \pm 0.8$ & $2.3 \pm 0.5$ & *8.8 $\pm$ 6.6 & *1.9 $\pm$ 1.4 \\
  & 10:21:16.5 & 11:12:27.9 & 20.6 & *2.4 $\pm$ 0.8 & *1.4 $\pm$ 0.5 & $21.9 \pm 6.6$ & $6.9 \pm 2.1$ \\ \hline
ELAN0101 & 01:01:16.5 & +02:01:57.4 & 18.1 & $5.1 \pm 0.8$ & $4.3 \pm 0.7$ & $42.8 \pm 7.8$ & $28.0 \pm 5.1$ \\
         & 01:01:16.9 & +02:01:49.9 & 21.7 & *1.6 $\pm$ 0.8 & *0.9 $\pm$ 0.5 & *-1.0 $\pm$ 7.9 & *-0.2 $\pm$ 1.7 \\\hline
J0119 & 01:19:07.5 & +02:05:58.3 & 20.2 & *2.2 $\pm$ 1.2 & *1.2 $\pm$ 0.6 & *7.5 $\pm$ 11.8 & *0.7 $\pm$ 1.1 \\
      & 01:19:05.4 & +02:05:12.8 & 20.6 & $3.5 \pm 1.2$ & $2.0 \pm 0.7$ & *3.5 $\pm$ 12.1 & *0.3 $\pm$ 1.2 \\ \hline
J1622 & 16:22:10.1 & +07:02:15.4 & 17.2 & \multirow{2}{*}{$8.7 \pm 0.9$}\tablefootmark{a} & \multirow{2}{*}{$6.9 \pm 0.7$} & \multirow{2}{*}{$28.6 \pm 8.4$} & \multirow{2}{*}{$7.2 \pm 2.1$} \\
      & 16:22:09.8 & +07:02:11.6 & 20.4 &  &  &  & \\ \hline

\hline
\end{tabular}
\tablefoot{
\tablefoottext{a}{The detection is consistent with both quasars.}
}
\end{minipage}
\end{center}

\FloatBarrier
\newpage\phantom{xyz}\newpage
\section{Interloper and non-SMG candidates}
\label{app:interloper}

In this appendix we list all the interloper and non-SMG candidates identified as described in Section~\ref{sec:extraction}. 

\begin{center}
\begin{minipage}{\textwidth}
\captionof{table}{Interloper and non-SMG candidates identified from the flux ratio between 850~$\mu$m and 450~$\mu$m detections.}
\label{tab:interlopers}
\centering 
\begin{tabular}{c c c c c c c c}
\hline\hline
ID & RA (J2000) & Dec (J2000) & $\Delta(\alpha,\delta)$ [\arcsec] & $f_{850}$ [mJy] & $f_{\rm 850, db}$ [mJy] & $f_{450}$ [mJy] & $f_{\rm 450, db}$ [mJy]\\    
\hline
1-IL.1 & 00:01:48.5 & +07:05:13.8 & 1.36 & 5.4 $\pm$ 1.7 & 3.1 $\pm$ 1.0 & 69.8 $\pm$ 22.9 & 7.6 $\pm$ 2.5\\ \hline
2-IL.2 & 00:18:04.3 & +16:06:56.8 & - & 24.3 $\pm$ 4.9 & - & 108.7 $\pm$ 24.0 & - \\ \hline
3-IL.1 & 02:39:56.3 & -00:41:57.8 & 0.51 & $14.6 \pm 1.4$ & $13.1 \pm 1.2$ & $62.0 \pm 14.0$ & 28.6 $\pm$ 6.5\\ \hline
4-IL.1 & 02:44:40.9 & -00:22:48.4 & 0.60 & 9.4 $\pm$ 1.4 & 8.1 $\pm$ 1.2 & 45.4 $\pm$ 14.3 & 3.0 $\pm$ 1.0\\ \hline
5-IL.1 & 10:13:01.7 & +03:36:10.8 & 0.6 & $11.1 \pm 1.1$ & $9.5 \pm 1.0$ & $28.6 \pm 9.1$ & 8.1 $\pm$ 2.6 \\
5-IL.2 & 10:12:50.2 & +03:34:26.8 & 0.6 & $9.4 \pm 1.0$ & $8.0 \pm 0.9$ & $36.6 \pm 8.2$ & 21.3 $\pm$ 4.8 \\
5-IL.3 & 10:12:34.0 & +03:33:42.8 & 0.9 & $9.0 \pm 1.9$ & $5.6 \pm 1.2$ & $53.2 \pm 16.0$ & 16.5 $\pm$ 4.9 \\
5-IL.4 & 10:12:37.9 & +03:32:20.8 & 0.9 & $8.0 \pm 1.7$ & $4.9 \pm 1.1$ & $57.5 \pm 15.0$ & 24.6 $\pm$ 6.4 \\
5-IL.5 & 10:12:47.4 & +03:35:32.8 & 1.0 & $4.6 \pm 1.1$ & $2.7 \pm 0.6$ & $28.6 \pm 8.2$ & 9.6 $\pm$ 2.8 \\
5-IL.6 & 10:12:49.5 & +03:37:22.8 & 1.1 & $4.7 \pm 1.2$ & $2.6 \pm 0.7$ & $42.3 \pm 8.9$ & 25.6 $\pm$ 5.4 \\
5-IL.7 & 10:12:41.6 & +03:35:12.8 & 1.2 & $5.2 \pm 1.5$ & $2.8 \pm 0.8$ & $32.3 \pm 10.0$ & 9.6 $\pm$ 2.9 \\
5-IL.8 & 10:12:55.3 & +03:35:54.8 & 1.2 & $3.2 \pm 0.9$ & $1.7 \pm 0.5$ & $21.6 \pm 7.1$ & 5.8 $\pm$ 1.9 \\ \hline
6-IL.1 & 10:21:14.2 & +11:12:51.9 & 0.6 & $7.2 \pm 0.9$ & $6.2 \pm 0.7$ & $28.6 \pm 6.8$ & 16.3 $\pm$ 3.9 \\
6-IL.2 & 10:21:34.0 & +11:13:55.9 & 0.7 & $9.5 \pm 1.3$ & $7.9 \pm 1.1$ & $54.4 \pm 11.4$ & 33.7 $\pm$ 7.1 \\
6-IL.3 & 10:21:12.0 & +11:17:19.9 & 0.7 & $10.4 \pm 1.7$ & $8.3 \pm 1.3$ & $54.2 \pm 12.4$ & 31.5 $\pm$ 7.2 \\
6-IL.4 & 10:21:26.1 & +11:12:31.9 & 0.8 & $6.2 \pm 1.0$ & $4.9 \pm 0.8$ & $30.2 \pm 9.3$ & 9.0 $\pm$ 2.8 \\
6-IL.5 & 10:21:12.0 & +11:11:53.9 & 0.8 & $4.6 \pm 0.8$ & $3.5 \pm 0.6$ & $21.9 \pm 7.2$ & 6.0 $\pm$ 2.0 \\
6-IL.6 & 10:21:21.4 & +11:10:27.9 & 0.8 & $4.7 \pm 1.0$ & $3.4 \pm 0.7$ & $31.3 \pm 8.6$ & 11.8 $\pm$ 3.2 \\
6-IL.7 & 10:21:33.7 & +11:14:19.9 & 1.0 & $5.6 \pm 1.4$ & $3.6 \pm 0.9$ & $49.4 \pm 12.2$ & 24.9 $\pm$ 6.2 \\
6-IL.8 & 10:21:25.6 & +11:08:05.9 & 1.2 & $4.2 \pm 1.2$ & $2.6 \pm 0.7$ & $49.5 \pm 11.6$ & 28.4 $\pm$ 6.7 \\ \hline
ELAN0101-IL.1 & 01:01:07.5 & +02:01:35.4 & 0.6 & $8.9 \pm 1.0$ & $8.1 \pm 0.9$ & $30.5 \pm 9.7$ & $8.2 \pm 2.6$ \\
ELAN0101-IL.2 & 01:01:16.9 & +02:02:21.4 & 0.7 & $5.4 \pm 0.9$ & $4.6 \pm 0.7$ & $25.3 \pm 7.8$ & $7.0 \pm 2.2$ \\ \hline
J0119-IL.1 & 01:18:48.8 & +02:03:08.2 & 1.3 & $7.3 \pm 2.4$ & $4.2 \pm 1.4$ & $76.8 \pm 24.3$ & 10.0 $\pm$ 3.2 \\ \hline
J1622-IL.1 & 16:22:11.8 & +07:02:13.5 & 0.7 & $8.0 \pm 1.0$ & $6.0 \pm 0.7$ & $27.6 \pm 8.7$ & 6.3 $\pm$ 2.0 \\
J1622-IL.2 & 16:22:12.8 & +07:05:19.5 & 0.9 & $7.0 \pm 1.6$ & $3.8 \pm 0.8$ & $44.9 \pm 13.5$ & 10.9 $\pm$ 3.3 \\
J1622-IL.3 & 16:21:57.5 & +06:59:29.5 & 1.3 & $4.3 \pm 1.3$ & $1.9 \pm 0.6$ & $50.2 \pm 13.2$ & 15.1 $\pm$ 3.9 \\ \hline
\hline
\end{tabular}
\end{minipage}
\end{center}

\FloatBarrier
\newpage\phantom{xyz}\newpage
\section{SMG candidates}
\label{app:smgcats}

In this appendix we report all the source catalogs for the nine fields in this study, built following the procedure
discussed in Section~\ref{sec:extraction}. Flux boosting and positional uncertainty are computed as described in Section~\ref{sec:diffnum}. 
All catalogues are ordered following the S/N of the sources.

\begin{center}
\begin{minipage}{\textwidth}
\captionof{table}{Position and fluxes of SMG candidates in the field of quasar pair 1. For non-detections at 450~$\mu$m, we report forced fluxes at the coordinate of the source and mark them with an asterisk.}
\label{tab:smg1}
\centering 
\begin{tabular}{c c c c c c c c}
\hline\hline
ID & RA (J2000) & Dec (J2000) & $\Delta(\alpha,\delta)$ [\arcsec] & $f_{850}$ [mJy] & $f_{\rm 850, db}$ [mJy] & $f_{450}$ [mJy] & $f_{\rm 450, db}$ [mJy]\\    
\hline
1-SMG.1 & 00:01:46.6 & +07:06:33.8 & 0.67 & 9.6 $\pm$ 1.4 & 7.8 $\pm$ 1.1 & *43.7 $\pm$ 20.4 & *3.9 $\pm$ 1.8\\ \hline
1-SMG.2 & 00:01:51.7 & +07:07:05.8 & 0.83 & 8.3 $\pm$ 1.6 & 6.2 $\pm$ 1.2 & *-11.2 $\pm$ 23.9 & *-1.0 $\pm$ 2.2\\ \hline
1-SMG.3 & 00:01:52.7 & +07:08:31.8 & 0.83 & 9.3 $\pm$ 1.8 & 6.9 $\pm$ 1.4 & *9.5 $\pm$ 20.1 & *0.9 $\pm$ 1.8\\ \hline
1-SMG.4 & 00:01:33.5 & +07:05:39.8 & 0.87 & 10.0 $\pm$ 2.1 & 7.1 $\pm$ 1.5 & *-24.1 $\pm$ 23.1 & *-2.2 $\pm$ 2.1\\ \hline
1-SMG.5 & 00:01:21.1 & +07:10:07.8 & 0.89 & 9.6 $\pm$ 2.1 & 6.7 $\pm$ 1.5 & *-3.2 $\pm$ 20.0 & *-0.3 $\pm$ 1.8\\ \hline
1-SMG.6 & 00:01:36.0 & +07:08:27.8 & 0.91 & 5.5 $\pm$ 1.2 & 3.8 $\pm$ 0.9 & *22.9 $\pm$ 14.6 & *2.1 $\pm$ 1.3\\ \hline
1-SMG.7 & 00:01:37.1 & +07:04:41.8 & 0.92 & 8.5 $\pm$ 1.9 & 5.9 $\pm$ 1.3 & *-21.1 $\pm$ 26.2 & *-1.9 $\pm$ 2.4\\ \hline
1-SMG.8 & 00:01:29.7 & +07:07:45.8 & 0.94 & 7.0 $\pm$ 1.6 & 4.7 $\pm$ 1.1 & *13.7 $\pm$ 23.3 & *1.2 $\pm$ 2.1\\ \hline
1-SMG.9 & 00:01:33.7 & +07:11:01.8 & 0.98 & 7.3 $\pm$ 1.8 & 4.8 $\pm$ 1.2 & *1.2 $\pm$ 17.1 & *0.1 $\pm$ 1.5\\ \hline
1-SMG.10 & 00:01:32.1 & +07:09:25.8 & 0.99 & 5.8 $\pm$ 1.4 & 3.8 $\pm$ 0.9 & *18.1 $\pm$ 15.0 & *1.6 $\pm$ 1.4\\ \hline
1-SMG.11 & 00:02:03.8 & +07:08:39.8 & 1.01 & 9.2 $\pm$ 2.3 & 6.0 $\pm$ 1.5 & *-30.8 $\pm$ 34.8 & *-2.8 $\pm$ 3.1\\ \hline
1-SMG.12 & 00:01:53.1 & +07:07:55.8 & 1.02 & 7.3 $\pm$ 1.8 & 4.8 $\pm$ 1.2 & *9.5 $\pm$ 22.2 & *0.9 $\pm$ 2.0\\ \hline
1-SMG.13 & 00:01:37.6 & +07:10:11.8 & 1.02 & 5.1 $\pm$ 1.3 & 3.3 $\pm$ 0.8 & *5.3 $\pm$ 13.3 & *0.5 $\pm$ 1.2\\ \hline
1-SMG.14 & 00:01:44.9 & +07:12:59.8 & 1.02 & 10.3 $\pm$ 2.6 & 6.7 $\pm$ 1.7 & *-6.3 $\pm$ 19.8 & *-0.6 $\pm$ 1.8\\ \hline
\hline
\end{tabular}
\end{minipage}
\end{center}

\begin{center}
\begin{minipage}{\textwidth}
\captionof{table}{Position and fluxes of SMG candidates in the field of quasar pair 2. For non-detections at 450~$\mu$m, we report forced fluxes at the coordinate of the source and mark them with an asterisk.}
\label{tab:smg2}
\centering 
\begin{tabular}{c c c c c c c c}
\hline\hline
ID & RA (J2000) & Dec (J2000) & $\Delta(\alpha,\delta)$ [\arcsec] & $f_{850}$ [mJy] & $f_{\rm 850, db}$ [mJy] & $f_{450}$ [mJy] & $f_{\rm 450, db}$ [mJy]\\    
\hline
2-SMG.1 & 00:18:11.4 & +16:10:40.8 & - & 12.5 $\pm$ 3.1 & - & *14.3 $\pm$ 18.87 & -\\ \hline
\hline
\end{tabular}
\end{minipage}
\end{center}

\begin{table*}
\caption{Position and fluxes of SMG candidates in the field of quasar pair 3. For non-detections at 450~$\mu$m, we report forced fluxes at the coordinate of the source and mark them with an asterisk.}
\label{tab:smg3}
\centering 
\begin{tabular}{c c c c c c c c}
\hline\hline
ID & RA (J2000) & Dec (J2000) & $\Delta(\alpha,\delta)$ [\arcsec] & $f_{850}$ [mJy] & $f_{\rm 850, db}$ [mJy] & $f_{450}$ [mJy] & $f_{\rm 450, db}$ [mJy]\\    
\hline
3-SMG.1 & 02:39:58.7 & -00:37:25.8 & 0.61 & 11.5 $\pm$ 1.2 & 9.9 $\pm$ 1.1 & *14.1 $\pm$ 12.3 & *2.1 $\pm$ 1.9\\ \hline
3-SMG.2 & 02:40:10.3 & -00:43:09.8 & 0.56 & 13.0 $\pm$ 1.5 & 11.0 $\pm$ 1.2 & *40.9 $\pm$ 18.5 & *6.2 $\pm$ 2.8\\ \hline
3-SMG.3 & 02:39:45.5 & -00:36:19.8 & 0.68 & 13.7 $\pm$ 1.8 & 11.0 $\pm$ 1.4 & *-16.7 $\pm$ 21.9 & *-2.5 $\pm$ 3.3\\ \hline
3-SMG.4 & 02:39:50.2 & -00:43:17.8 & 0.70 & 11.0 $\pm$ 1.5 & 8.5 $\pm$ 1.2 & *18.1 $\pm$ 17.9 & *2.7 $\pm$ 2.7\\ \hline
3-SMG.5 & 02:40:09.2 & -00:37:47.8 & 0.71 & 7.9 $\pm$ 1.1 & 6.0 $\pm$ 0.9 & *23.9 $\pm$ 10.4 & *3.6 $\pm$ 1.6\\ \hline
3-SMG.6 & 02:39:59.0 & -00:40:05.8 & 0.71 & 7.2 $\pm$ 1.1 & 5.4 $\pm$ 0.8 & *15.6 $\pm$ 12.1 & *2.4 $\pm$ 1.8\\ \hline
3-SMG.7 & 02:40:03.2 & -00:39:35.8 & 0.76 & 5.6 $\pm$ 0.9 & 3.8 $\pm$ 0.6 & *5.9 $\pm$ 9.6 & *0.9 $\pm$ 1.4\\ \hline
3-SMG.8 & 02:39:53.9 & -00:38:37.8 & 0.88 & 5.6 $\pm$ 1.1 & 3.2 $\pm$ 0.6 & *-1.9 $\pm$ 11.9 & *-0.3 $\pm$ 1.8\\ \hline
3-SMG.9 & 02:39:57.1 & -00:34:41.8 & 0.89 & 8.0 $\pm$ 1.7 & 4.5 $\pm$ 0.9 & *-4.3 $\pm$ 18.8 & *-0.6 $\pm$ 2.8\\ \hline
3-SMG.10 & 02:40:14.8 & -00:36:35.8 & 0.91 & 9.9 $\pm$ 2.1 & 5.5 $\pm$ 1.1 & *26.1 $\pm$ 13.9 & *3.9 $\pm$ 2.1\\ \hline
3-SMG.11 & 02:39:56.8 & -00:37:47.8 & 0.93 & 5.8 $\pm$ 1.3 & 3.1 $\pm$ 0.7 & *4.5 $\pm$ 11.5 & *0.7 $\pm$ 1.7\\ \hline
3-SMG.12 & 02:40:14.4 & -00:39:17.8 & 0.93 & 5.6 $\pm$ 1.2 & 3.0 $\pm$ 0.6 & *-2.2 $\pm$ 11.6 & v-0.3 $\pm$ 1.7\\ \hline
3-SMG.13 & 02:39:54.3 & -00:39:21.8 & 0.94 & 5.1 $\pm$ 1.1 & 2.7 $\pm$ 0.6 & *3.9 $\pm$ 13.1 & *0.6 $\pm$ 2.0\\ \hline
3-SMG.14 & 02:39:58.7 & -00:40:49.8 & 0.94 & 5.5 $\pm$ 1.2 & 2.9 $\pm$ 0.6 & *-4.5 $\pm$ 13.1 & *-0.7 $\pm$ 2.0\\ \hline
3-SMG.15 & 02:40:07.2 & -00:34:07.8 & 0.94 & 8.1 $\pm$ 1.8 & 4.3 $\pm$ 0.9 & *-7.7 $\pm$ 17.2 & *-1.2 $\pm$ 2.6\\ \hline
3-SMG.16 & 02:40:08.8 & -00:35:09.8 & 0.94 & 6.4 $\pm$ 1.4 & 3.3 $\pm$ 0.7 & *2.4 $\pm$ 12.5 & *0.4 $\pm$ 1.9\\ \hline
3-SMG.17 & 02:40:09.8 & -00:43:17.8 & 0.96 & 6.3 $\pm$ 1.5 & 3.3 $\pm$ 0.8 & *-2.1 $\pm$ 18.7 & *-0.3 $\pm$ 2.8\\ \hline
3-SMG.18 & 02:40:07.9 & -00:37:33.8 & 0.96 & 4.7 $\pm$ 1.1 & 2.4 $\pm$ 0.6 & *-2.0 $\pm$ 10.4 & *-0.3 $\pm$ 1.6\\ \hline
3-SMG.19 & 02:39:48.4 & -00:36:51.8 & 0.96 & 6.3 $\pm$ 1.5 & 3.2 $\pm$ 0.7 & *3.5 $\pm$ 15.2 & *0.5 $\pm$ 2.3\\ \hline
3-SMG.20 & 02:39:43.9 & -00:37:33.8 & 1.01 & 7.4 $\pm$ 1.8 & 3.7 $\pm$ 0.9 & *-0.5 $\pm$ 19.8 & *-0.1 $\pm$ 3.0\\ \hline
3-SMG.21 & 02:40:10.0 & -00:35:07.8 & 1.01 & 6.2 $\pm$ 1.5 & 3.1 $\pm$ 0.8 & *19.6 $\pm$ 12.6 & *3.0 $\pm$ 1.9\\ \hline
3-SMG.22 & 02:40:14.4 & -00:40:55.8 & 1.02 & 6.5 $\pm$ 1.6 & 3.2 $\pm$ 0.8 & *19.1 $\pm$ 13.5 & *2.9 $\pm$ 2.0\\ \hline
3-SMG.23 & 02:40:08.0 & -00:39:35.8 & 1.05 & 3.9 $\pm$ 1.0 & 1.9 $\pm$ 0.5 & *-7.7 $\pm$ 9.6 & *-1.2 $\pm$ 1.5\\ \hline
3-SMG.24 & 02:40:17.8 & -00:43:31.8 & 1.05 & 7.2 $\pm$ 1.8 & 3.5 $\pm$ 0.9 & *-16.2 $\pm$ 23.3 & *-2.4 $\pm$ 3.5\\ \hline
3-SMG.25 & 02:40:07.9 & -00:41:45.8 & 1.05 & 4.9 $\pm$ 1.2 & 2.4 $\pm$ 0.6 & *-16.5 $\pm$ 15.2 & *-2.5 $\pm$ 2.3\\ \hline
\hline
\end{tabular}
\end{table*}

\begin{table*}
\caption{Position and fluxes of SMG candidates in the field of quasar pair 4. For non-detections at 450~$\mu$m, we report forced fluxes at the coordinate of the source and mark them with an asterisk.}
\label{tab:smg4}
\centering 
\begin{tabular}{c c c c c c c c}
\hline\hline
ID & RA (J2000) & Dec (J2000) & $\Delta(\alpha,\delta)$ [\arcsec] & $f_{850}$ [mJy] & $f_{\rm 850, db}$ [mJy] & $f_{450}$ [mJy] & $f_{\rm 450, db}$ [mJy]\\    
\hline
4-SMG.1 & 02:44:50.2 & -00:27:10.4 & 0.55 & 13.1 $\pm$ 1.7 & 11.4 $\pm$ 1.5 & *-0.6 $\pm$ 23.0 & *-0.0 $\pm$ 1.4\\ \hline
4-SMG.2 & 02:44:26.5 & -00:19:54.4 & 0.65 & 15.4 $\pm$ 2.4 & 13.0 $\pm$ 2.0 & *32.4 $\pm$ 22.5 & *1.9 $\pm$ 1.3\\ \hline
4-SMG.3 & 02:44:33.4 & -00:20:50.4 & 0.67 & 13.4 $\pm$ 2.1 & 11.2 $\pm$ 1.8 & *16.2 $\pm$ 24.1 & *1.0 $\pm$ 1.4\\ \hline
4-SMG.4 & 02:44:49.7 & -00:23:54.4 & 0.70 & 9.3 $\pm$ 1.6 & 7.6 $\pm$ 1.3 & *-17.3 $\pm$ 18.2 & *-1.0 $\pm$ 1.1\\ \hline
4-SMG.5 & 02:44:45.9 & -00:22:50.4 & 0.75 & 7.3 $\pm$ 1.4 & 5.8 $\pm$ 1.1 & *-16.8 $\pm$ 15.7 & *-1.0 $\pm$ 0.9\\ \hline
4-SMG.6 & 02:44:47.7 & -00:22:46.4 & 0.82 & 7.1 $\pm$ 1.4 & 5.2 $\pm$ 1.1 & *13.4 $\pm$ 17.1 & *0.8 $\pm$ 1.0\\ \hline
4-SMG.7 & 02:44:36.1 & -00:22:48.4 & 0.82 & 7.6 $\pm$ 1.6 & 5.6 $\pm$ 1.2 & *-33.8 $\pm$ 16.2 & *-2.0 $\pm$ 1.0\\ \hline
4-SMG.8 & 02:44:42.9 & -00:22:22.4 & 0.83 & 6.7 $\pm$ 1.4 & 4.8 $\pm$ 1.0 & *-7.5 $\pm$ 14.8 & *-0.5 $\pm$ 0.9\\ \hline
4-SMG.9 & 02:44:41.9 & -00:25:14.4 & 0.86 & 6.5 $\pm$ 1.4 & 4.6 $\pm$ 1.0 & *14.3 $\pm$ 15.8 & *0.9 $\pm$ 0.9\\ \hline
4-SMG.10 & 02:44:57.1 & -00:23:30.4 & 0.87 & 7.8 $\pm$ 1.7 & 5.5 $\pm$ 1.2 & *1.0 $\pm$ 20.1 & *0.1 $\pm$ 1.2\\ \hline
4-SMG.11 & 02:44:45.8 & -00:23:48.4 & 0.88 & 5.8 $\pm$ 1.3 & 4.0 $\pm$ 0.9 & *29.7 $\pm$ 15.3 & *1.8 $\pm$ 0.9\\ \hline
4-SMG.12 & 02:44:44.9 & -00:24:56.4 & 0.89 & 6.0 $\pm$ 1.4 & 4.1 $\pm$ 1.0 & *22.7 $\pm$ 15.9 & *1.4 $\pm$ 1.0\\ \hline
4-SMG.13 & 02:45:02.1 & -00:24:48.4 & 0.89 & 8.8 $\pm$ 2.1 & 6.1 $\pm$ 1.4 & *21.2 $\pm$ 23.8 & *1.3 $\pm$ 1.4\\ \hline
4-SMG.14 & 02:44:40.1 & -00:20:20.4 & 0.90 & 7.7 $\pm$ 1.8 & 5.3 $\pm$ 1.2 & *12.5 $\pm$ 18.2 & *0.8 $\pm$ 1.1\\ \hline
4-SMG.15 & 02:44:33.9 & -00:18:48.4 & 0.92 & 10.3 $\pm$ 2.5 & 7.0 $\pm$ 1.7 & *9.3 $\pm$ 22.8 & *0.6 $\pm$ 1.4\\ \hline
4-SMG.16 & 02:44:29.4 & -00:24:36.4 & 0.93 & 7.2 $\pm$ 1.7 & 4.9 $\pm$ 1.2 & *-39.2 $\pm$ 21.4 & *-2.3 $\pm$ 1.3\\ \hline
4-SMG.17 & 02:44:49.8 & -00:23:36.4 & 0.95 & 6.3 $\pm$ 1.5 & 4.2 $\pm$ 1.0 & *-2.1 $\pm$ 18.2 & *-0.1 $\pm$ 1.1\\ \hline
4-SMG.18 & 02:44:54.3 & -00:19:58.4 & 0.99 & 10.7 $\pm$ 2.7 & 7.0 $\pm$ 1.8 & *-16.1 $\pm$ 31.0 & *-1.0 $\pm$ 1.9\\ \hline
\hline
\end{tabular}
\end{table*}

\begin{table*}
\caption{Position and fluxes of SMG candidates in the field of quasar pair 5. For non-detections at 450~$\mu$m, we report forced fluxes at the coordinate of the source and mark them with an asterisk.}
\label{tab:smg5}
\centering 
\begin{tabular}{c c c c c c c c}
\hline\hline
ID & RA (J2000) & Dec (J2000) & $\Delta(\alpha,\delta)$ [\arcsec] & $f_{850}$ [mJy] & $f_{\rm 850, db}$ [mJy] & $f_{450}$ [mJy] & $f_{\rm 450, db}$ [mJy]\\    
\hline
5-SMG.1 & 10:12:52.7 & +03:36:22.8 & 0.62 & 8.4 $\pm$ 1.0 & 6.9 $\pm$ 0.8 & *6.6 $\pm$ 7.3 & *1.4 $\pm$ 1.6\\ \hline
5-SMG.2 & 10:12:47.7 & +03:34:48.8 & 0.67 & 8.3 $\pm$ 1.1 & 6.5 $\pm$ 0.8 & 38.9 $\pm$ 8.4 & 23.1 $\pm$ 5.0\\ \hline
5-SMG.3 & 10:12:46.2 & +03:36:24.8 & 0.71 & 8.7 $\pm$ 1.3 & 6.6 $\pm$ 1.0 & *7.3 $\pm$ 9.4 & *1.6 $\pm$ 2.0\\ \hline
5-SMG.4 & 10:12:48.7 & +03:38:08.8 & 0.71 & 9.0 $\pm$ 1.3 & 6.8 $\pm$ 1.0 & *9.1 $\pm$ 9.5 & *1.9 $\pm$ 2.0\\ \hline
5-SMG.5 & 10:12:41.0 & +03:30:56.8 & 0.72 & 12.1 $\pm$ 1.9 & 9.2 $\pm$ 1.4 & *33.9 $\pm$ 16.6 & *7.2 $\pm$ 3.5\\ \hline
5-SMG.6 & 10:12:44.5 & +03:32:02.8 & 0.75 & 8.5 $\pm$ 1.4 & 6.1 $\pm$ 1.0 & *18.3 $\pm$ 13.4 & *3.9 $\pm$ 2.9\\ \hline
5-SMG.7 & 10:12:53.0 & +03:35:36.8 & 0.75 & 5.4 $\pm$ 0.9 & 3.9 $\pm$ 0.7 & *16.0 $\pm$ 7.2 & *3.4 $\pm$ 1.5\\ \hline
5-SMG.8 & 10:13:00.1 & +03:32:34.8 & 0.76 & 6.9 $\pm$ 1.2 & 4.9 $\pm$ 0.8 & *15.4 $\pm$ 10.6 & *3.3 $\pm$ 2.3\\ \hline
5-SMG.9 & 10:12:46.9 & +03:37:22.8 & 0.78 & 7.4 $\pm$ 1.3 & 5.1 $\pm$ 0.9 & *24.6 $\pm$ 9.9 & *5.3 $\pm$ 2.1\\ \hline
5-SMG.10 & 10:13:04.5 & +03:34:28.8 & 0.80 & 7.2 $\pm$ 1.4 & 4.8 $\pm$ 0.9 & *7.1 $\pm$ 11.5 & *1.5 $\pm$ 2.5\\ \hline
5-SMG.11 & 10:12:51.9 & +03:36:16.8 & 0.81 & 5.1 $\pm$ 1.0 & 3.4 $\pm$ 0.6 & *9.5 $\pm$ 7.4 & *2.0 $\pm$ 1.6\\ \hline
5-SMG.12 & 10:12:50.5 & +03:35:58.8 & 0.80 & 5.2 $\pm$ 1.0 & 3.5 $\pm$ 0.7 & *8.9 $\pm$ 7.5 & *1.9 $\pm$ 1.6\\ \hline
5-SMG.13 & 10:13:15.0 & +03:33:34.8 & 0.85 & 8.7 $\pm$ 1.8 & 5.6 $\pm$ 1.1 & *-13.0 $\pm$ 15.2 & *-2.8 $\pm$ 3.3\\ \hline
5-SMG.14 & 10:12:57.5 & +03:40:48.8 & 0.89 & 8.9 $\pm$ 1.9 & 5.6 $\pm$ 1.2 & *-10.1 $\pm$ 13.6 & *-2.2 $\pm$ 2.9\\ \hline
5-SMG.15 & 10:12:56.3 & +03:34:24.8 & 0.92 & 4.5 $\pm$ 1.0 & 2.7 $\pm$ 0.6 & *-1.7 $\pm$ 8.0 & *-0.4 $\pm$ 1.7\\ \hline
5-SMG.16 & 10:12:53.5 & +03:34:38.8 & 0.92 & 4.3 $\pm$ 0.9 & 2.6 $\pm$ 0.6 & *20.2 $\pm$ 7.6 & *4.6 $\pm$ 1.7\\ \hline
5-SMG.17 & 10:13:01.2 & +03:34:30.8 & 0.93 & 5.2 $\pm$ 1.1 & 3.1 $\pm$ 0.7 & *0.5 $\pm$ 9.9 & *0.1 $\pm$ 2.1\\ \hline
5-SMG.18 & 10:13:03.2 & +03:32:28.8 & 0.93 & 5.4 $\pm$ 1.2 & 3.3 $\pm$ 0.7 & *0.9 $\pm$ 11.2 & *0.2 $\pm$ 2.4\\ \hline
5-SMG.19 & 10:13:08.9 & +03:35:40.8 & 0.95 & 6.3 $\pm$ 1.4 & 3.8 $\pm$ 0.9 & *13.0 $\pm$ 11.5 & *2.8 $\pm$ 2.4\\ \hline
5-SMG.20 & 10:13:00.2 & +03:31:54.8 & 0.96 & 5.2 $\pm$ 1.2 & 3.1 $\pm$ 0.7 & *7.3 $\pm$ 11.1 & *1.6 $\pm$ 2.4\\ \hline
5-SMG.21 & 10:12:48.6 & +03:34:04.8 & 0.96 & 4.8 $\pm$ 1.1 & 2.8 $\pm$ 0.6 & *8.4 $\pm$ 9.2 & *1.8 $\pm$ 2.0\\ \hline
5-SMG.22 & 10:12:53.4 & +03:37:22.8 & 0.96 & 5.0 $\pm$ 1.1 & 2.9 $\pm$ 0.7 & *3.9 $\pm$ 8.2 & *0.8 $\pm$ 1.7\\ \hline
5-SMG.23 & 10:13:06.1 & +03:38:52.8 & 0.98 & 6.8 $\pm$ 1.6 & 4.0 $\pm$ 0.9 & *-10.2 $\pm$ 12.4 & *-2.2 $\pm$ 2.6\\ \hline
5-SMG.24 & 10:12:44.3 & +03:38:12.8 & 1.00 & 5.9 $\pm$ 1.4 & 3.4 $\pm$ 0.8 & *-9.0 $\pm$ 10.8 & *-1.9 $\pm$ 2.3\\ \hline
5-SMG.25 & 10:12:48.6 & +03:35:36.8 & 1.01 & 4.2 $\pm$ 1.0 & 2.4 $\pm$ 0.6 & *9.6 $\pm$ 7.9 & *2.0 $\pm$ 1.7\\ \hline
5-SMG.26 & 10:12:58.6 & +03:35:42.8 & 1.01 & 4.0 $\pm$ 1.0 & 2.3 $\pm$ 0.6 & *1.4 $\pm$ 7.9 & *0.3 $\pm$ 1.7\\ \hline
5-SMG.27 & 10:12:48.5 & +03:38:46.8 & 1.02 & 5.8 $\pm$ 1.4 & 3.3 $\pm$ 0.8 & *-9.7 $\pm$ 9.7 & *-2.1 $\pm$ 2.1\\ \hline
5-SMG.28 & 10:12:53.0 & +03:36:30.8 & 1.03 & 4.0 $\pm$ 1.0 & 2.3 $\pm$ 0.6 & *-5.1 $\pm$ 7.4 & *-1.1 $\pm$ 1.6\\ \hline
5-SMG.29 & 10:13:00.1 & +03:36:34.8 & 1.04 & 4.4 $\pm$ 1.1 & 2.5 $\pm$ 0.6 & *-7.1 $\pm$ 8.6 & *-1.5 $\pm$ 1.8\\ \hline
5-SMG.30 & 10:12:40.3 & +03:38:06.8 & 1.00 & 6.4 $\pm$ 1.6 & 3.7 $\pm$ 0.9 & *18.9 $\pm$ 11.3 & *4.0 $\pm$ 2.4\\ \hline
\hline
\end{tabular}
\end{table*}

\begin{table*}
\caption{Position and fluxes of SMG candidates in the field of quasar pair 6. For non-detections at 450~$\mu$m, we report forced fluxes at the coordinate of the source and mark them with an asterisk.}
\label{tab:smg6}
\centering 
\begin{tabular}{c c c c c c c c}
\hline\hline
ID & RA (J2000) & Dec (J2000) & $\Delta(\alpha,\delta)$ [\arcsec] & $f_{850}$ [mJy] & $f_{\rm 850, db}$ [mJy] & $f_{450}$ [mJy] & $f_{\rm 450, db}$ [mJy]\\    
\hline
6-SMG.1 & 10:21:00.2 & +11:10:01.9 & 0.58 & 11.3 $\pm$ 1.3 & 9.8 $\pm$ 1.1 & *-5.2 $\pm$ 11.7 & *-1.1 $\pm$ 2.5\\ \hline
6-SMG.2 & 10:21:25.4 & +11:09:41.9 & 0.72 & 7.0 $\pm$ 1.1 & 5.7 $\pm$ 0.9 & *-1.0 $\pm$ 11.1 & *-0.2 $\pm$ 2.4\\ \hline
6-SMG.3 & 10:21:22.0 & +11:10:31.9 & 0.73 & 6.4 $\pm$ 1.0 & 5.1 $\pm$ 0.8 & *9.6 $\pm$ 9.0 & *2.1 $\pm$ 1.9\\ \hline
6-SMG.4 & 10:21:08.3 & +11:13:09.9 & 0.73 & 7.3 $\pm$ 1.2 & 5.9 $\pm$ 0.9 & *4.5 $\pm$ 8.6 & *1.0 $\pm$ 1.9\\ \hline
6-SMG.5 & 10:21:08.6 & +11:12:35.9 & 0.75 & 6.2 $\pm$ 1.0 & 4.9 $\pm$ 0.8 & *13.0 $\pm$ 7.9 & *2.8 $\pm$ 1.7\\ \hline
6-SMG.6 & 10:21:20.3 & +11:16:59.9 & 0.77 & 9.5 $\pm$ 1.7 & 7.4 $\pm$ 1.3 & *14.7 $\pm$ 13.5 & *3.2 $\pm$ 2.9\\ \hline
6-SMG.7 & 10:21:07.6 & +11:14:21.9 & 0.78 & 7.4 $\pm$ 1.3 & 5.7 $\pm$ 1.0 & *7.1 $\pm$ 11.0 & *1.5 $\pm$ 2.4\\ \hline
6-SMG.8 & 10:21:15.7 & +11:13:17.9 & 0.82 & 4.4 $\pm$ 0.9 & 3.3 $\pm$ 0.6 & *-6.1 $\pm$ 6.9 & *-1.3 $\pm$ 1.5\\ \hline
6-SMG.9 & 10:21:19.5 & +11:06:53.9 & 0.84 & 7.6 $\pm$ 1.6 & 5.5 $\pm$ 1.1 & *0.1 $\pm$ 15.7 & *0.0 $\pm$ 3.4\\ \hline
6-SMG.10 & 10:21:12.5 & +11:07:29.9 & 0.85 & 6.7 $\pm$ 1.4 & 4.8 $\pm$ 1.0 & *0.5 $\pm$ 13.3 & *0.1 $\pm$ 2.9\\ \hline
6-SMG.11 & 10:21:08.7 & +11:12:07.9 & 0.86 & 4.5 $\pm$ 1.0 & 3.2 $\pm$ 0.7 & *-2.6 $\pm$ 7.8 & *-0.6 $\pm$ 1.7\\ \hline
6-SMG.12 & 10:21:20.0 & +11:15:51.9 & 0.86 & 7.2 $\pm$ 1.6 & 5.1 $\pm$ 1.1 & *-14.4 $\pm$ 10.6 & *-3.1 $\pm$ 2.3\\ \hline
6-SMG.13 & 10:21:09.3 & +11:12:23.9 & 0.87 & 4.4 $\pm$ 1.0 & 3.1 $\pm$ 0.7 & *3.2 $\pm$ 7.6 & *0.7 $\pm$ 1.6\\ \hline
6-SMG.14 & 10:21:15.2 & +11:14:47.9 & 0.88 & 5.0 $\pm$ 1.1 & 3.5 $\pm$ 0.8 & *4.2 $\pm$ 8.5 & *0.9 $\pm$ 1.8\\ \hline
6-SMG.15 & 10:21:24.2 & +11:12:21.9 & 0.88 & 4.4 $\pm$ 1.0 & 3.0 $\pm$ 0.7 & *10.5 $\pm$ 8.8 & *2.3 $\pm$ 1.9\\ \hline
6-SMG.16 & 10:21:36.2 & +11:13:33.9 & 0.88 & 6.3 $\pm$ 1.4 & 4.3 $\pm$ 1.0 & *19.8 $\pm$ 12.7 & *4.3 $\pm$ 2.7\\ \hline
6-SMG.17 & 10:21:10.4 & +11:11:11.9 & 0.89 & 4.0 $\pm$ 0.9 & 2.8 $\pm$ 0.6 & *-15.6 $\pm$ 8.0 & *-3.4 $\pm$ 1.7\\ \hline
6-SMG.18 & 10:20:53.2 & +11:11:53.8 & 0.89 & 7.6 $\pm$ 1.8 & 5.2 $\pm$ 1.2 & *-3.4 $\pm$ 15.0 & *-0.7 $\pm$ 3.2\\ \hline
6-SMG.19 & 10:21:07.6 & +11:14:35.9 & 0.90 & 5.5 $\pm$ 1.3 & 3.7 $\pm$ 0.9 & *5.9 $\pm$ 11.2 & *1.3 $\pm$ 2.4\\ \hline
6-SMG.20 & 10:21:18.2 & +11:07:11.9 & 0.90 & 5.8 $\pm$ 1.4 & 3.9 $\pm$ 0.9 & *4.9 $\pm$ 13.8 & *1.1 $\pm$ 3.0\\ \hline
6-SMG.21 & 10:21:10.6 & +11:14:23.9 & 0.91 & 4.9 $\pm$ 1.2 & 3.3 $\pm$ 0.8 & *-1.1 $\pm$ 9.4 & *-0.2 $\pm$ 2.0\\ \hline
6-SMG.22 & 10:21:05.6 & +11:14:49.9 & 0.94 & 5.5 $\pm$ 1.3 & 3.7 $\pm$ 0.9 & *24.9 $\pm$ 11.4 & *5.4 $\pm$ 2.5\\ \hline
6-SMG.23 & 10:21:06.0 & +11:10:57.9 & 0.94 & 4.5 $\pm$ 1.1 & 3.0 $\pm$ 0.7 & *-4.9 $\pm$ 9.6 & *-1.1 $\pm$ 2.1\\ \hline
6-SMG.24 & 10:21:06.0 & +11:15:37.9 & 0.95 & 5.7 $\pm$ 1.4 & 3.8 $\pm$ 0.9 & *3.1 $\pm$ 10.8 & *0.7 $\pm$ 2.3\\ \hline
6-SMG.25 & 10:21:33.7 & +11:09:05.9 & 0.95 & 5.5 $\pm$ 1.3 & 3.6 $\pm$ 0.9 & *-11.8 $\pm$ 11.6 & *-2.6 $\pm$ 2.5\\ \hline
6-SMG.26 & 10:20:59.5 & +11:10:45.9 & 0.95 & 5.2 $\pm$ 1.3 & 3.5 $\pm$ 0.8 & *20.9 $\pm$ 10.7 & *4.5 $\pm$ 2.3\\ \hline
6-SMG.27 & 10:21:25.2 & +11:13:51.9 & 0.97 & 4.4 $\pm$ 1.1 & 2.9 $\pm$ 0.7 & *27.7 $\pm$ 10.1 & *6.6 $\pm$ 2.4\\ \hline
6-SMG.28 & 10:20:59.6 & +11:12:49.9 & 1.21 & 4.6 $\pm$ 1.4 & 2.8 $\pm$ 0.8 & 31.6 $\pm$ 9.1 & 10.5 $\pm$ 3.1\\ \hline
\hline
\end{tabular}
\end{table*}

\begin{table*}
\caption{Position and fluxes of SMG candidates in the field of quasar pair ELAN0101. For non-detections at 450~$\mu$m, we report forced fluxes at the coordinate of the source and mark them with an asterisk.}
\label{tab:smg0101}
\centering 
\begin{tabular}{c c c c c c c c}
\hline\hline
ID & RA (J2000) & Dec (J2000) & $\Delta(\alpha,\delta)$ [\arcsec] & $f_{850}$ [mJy] & $f_{\rm 850, db}$ [mJy] & $f_{450}$ [mJy] & $f_{\rm 450, db}$ [mJy]\\    
\hline
ELAN0101-SMG.1 & 01:01:02.1 & +02:05:45.4 & 0.66 & 9.8 $\pm$ 1.5 & 8.4 $\pm$ 1.3 & *15.6 $\pm$ 12.2 & *3.3 $\pm$ 2.6\\ \hline
ELAN0101-SMG.2 & 01:01:24.7 & +02:02:25.4 & 0.74 & 5.9 $\pm$ 1.1 & 4.9 $\pm$ 0.9 & *23.4 $\pm$ 11.1 & *5.0 $\pm$ 2.4\\ \hline
ELAN0101-SMG.3 & 01:01:25.3 & +02:03:51.4 & 0.80 & 7.2 $\pm$ 1.4 & 5.7 $\pm$ 1.1 & *21.7 $\pm$ 12.9 & *4.6 $\pm$ 2.7\\ \hline
ELAN0101-SMG.4 & 01:01:17.3 & +02:06:23.4 & 0.81 & 7.9 $\pm$ 1.6 & 6.2 $\pm$ 1.3 & *14.2 $\pm$ 12.4 & *3.0 $\pm$ 2.6\\ \hline
ELAN0101-SMG.5 & 01:01:13.1 & +02:00:37.4 & 0.81 & 4.4 $\pm$ 0.9 & 3.4 $\pm$ 0.7 & *2.0 $\pm$ 8.5 & *0.4 $\pm$ 1.8 \\ \hline
ELAN0101-SMG.6 & 01:01:23.9 & +02:04:31.4 & 0.83 & 7.9 $\pm$ 1.6 & 6.1 $\pm$ 1.3 & *-10.4 $\pm$ 12.0 & *-2.2 $\pm$ 2.6\\ \hline
ELAN0101-SMG.7 & 01:01:04.1 & +02:00:27.4 & 0.84 & 5.9 $\pm$ 1.2 & 4.6 $\pm$ 1.0 & *8.9 $\pm$ 13.3 & *1.9 $\pm$ 2.8\\ \hline
ELAN0101-SMG.8 & 01:01:30.4 & +02:03:51.4 & 0.85 & 7.4 $\pm$ 1.6 & 5.7 $\pm$ 1.2 & *12.1 $\pm$ 14.6 & *2.6 $\pm$ 3.1\\ \hline
ELAN0101-SMG.9 & 01:01:18.5 & +01:59:29.4 & 0.86 & 4.6 $\pm$ 1.0 & 3.5 $\pm$ 0.8 & *11.2 $\pm$ 10.8 & *2.4 $\pm$ 2.3\\ \hline
ELAN0101-SMG.10 & 01:01:08.4 & +01:59:11.4 & 0.87 & 6.1 $\pm$ 1.4 & 4.6 $\pm$ 1.0 & *21.6 $\pm$ 12.4 & *4.6 $\pm$ 2.6\\ \hline
ELAN0101-SMG.11 & 01:01:10.0 & +01:59:45.4 & 0.89 & 4.9 $\pm$ 1.1 & 3.7 $\pm$ 0.8 & *13.5 $\pm$ 11.1 & *2.9 $\pm$ 2.4\\ \hline
ELAN0101-SMG.12 & 01:01:16.8 & +01:57:49.4 & 0.93 & 4.4 $\pm$ 1.1 & 3.3 $\pm$ 0.8 & *24.4 $\pm$ 12.2 & *5.2 $\pm$ 2.6\\ \hline
ELAN0101-SMG.13 & 01:01:11.9 & +02:00:17.4 & 0.93 & 3.9 $\pm$ 0.9 & 2.9 $\pm$ 0.7 & *11.3 $\pm$ 9.2 & *2.4 $\pm$ 1.9\\ \hline
ELAN0101-SMG.14 & 01:01:18.1 & +02:02:13.4 & 0.93 & 3.6 $\pm$ 0.9 & 2.7 $\pm$ 0.6 & *-5.0 $\pm$ 8.1 & *-1.1 $\pm$ 1.7\\ \hline
ELAN0101-SMG.15 & 01:00:55.2 & +02:00:25.4 & 0.94 & 7.2 $\pm$ 1.7 & 5.2 $\pm$ 1.3 & *7.5 $\pm$ 16.8 & *1.6 $\pm$ 3.6\\ \hline
ELAN0101-SMG.16 & 01:01:15.9 & +01:56:25.4 & 0.97 & 6.1 $\pm$ 1.5 & 4.4 $\pm$ 1.1 & *-15.1 $\pm$ 16.7 & *-3.2 $\pm$ 3.5\\ \hline
\hline
\end{tabular}
\end{table*}

\begin{table*}
\caption{Position and fluxes of SMG candidates in the field of quasar pair J0119. For non-detections at 450~$\mu$m, we report forced fluxes at the coordinate of the source and mark them with an asterisk.}
\label{tab:smg0119}
\centering 
\begin{tabular}{c c c c c c c c}
\hline\hline
ID & RA (J2000) & Dec (J2000) & $\Delta(\alpha,\delta)$ [\arcsec] & $f_{850}$ [mJy] & $f_{\rm 850, db}$ [mJy] & $f_{450}$ [mJy] & $f_{\rm 450, db}$ [mJy]\\    
\hline
J0119-SMG.1 & 01:19:16.3 & +02:01:50.2 & 0.67 & 10.1 $\pm$ 1.6 & 8.2 $\pm$ 1.3 & *-34.4 $\pm$ 20.1 & *-3.3 $\pm$ 2.0\\ \hline
J0119-SMG.2 & 01:19:09.0 & +02:00:12.2 & 0.74 & 13.0 $\pm$ 2.4 & 10.3 $\pm$ 1.9 & *22.3 $\pm$ 32.1 & *2.2 $\pm$ 3.1\\ \hline
J0119-SMG.3 & 01:19:12.3 & +02:03:10.2 & 0.78 & 7.3 $\pm$ 1.4 & 5.6 $\pm$ 1.1 & 55.9 $\pm$ 15.8 & 9.1 $\pm$ 2.6\\ \hline
J0119-SMG.4 & 01:19:12.8 & +02:05:44.2 & 0.81 & 6.5 $\pm$ 1.3 & 4.9 $\pm$ 1.0 & *-1.7 $\pm$ 14.2 & *-0.2 $\pm$ 1.4\\ \hline
J0119-SMG.5 & 01:19:18.4 & +02:07:46.2 & 0.82 & 8.4 $\pm$ 1.7 & 6.3 $\pm$ 1.3 & *9.2 $\pm$ 18.5 & *0.9 $\pm$ 1.8\\ \hline
J0119-SMG.6 & 01:18:52.5 & +02:03:18.2 & 0.88 & 8.3 $\pm$ 1.9 & 6.0 $\pm$ 1.3 & *-1.9 $\pm$ 21.2 & *-0.2 $\pm$ 2.1\\ \hline
J0119-SMG.7 & 01:18:49.7 & +02:08:58.2 & 0.88 & 10.0 $\pm$ 2.3 & 7.2 $\pm$ 1.6 & *-11.1 $\pm$ 18.7 & *-1.1 $\pm$ 1.8\\ \hline
J0119-SMG.8 & 01:19:18.8 & +02:08:12.2 & 0.88 & 8.3 $\pm$ 1.9 & 5.9 $\pm$ 1.4 & *-5.8 $\pm$ 20.3 & *-0.6 $\pm$ 2.0\\ \hline
J0119-SMG.9 & 01:19:23.9 & +02:01:52.2 & 0.90 & 9.0 $\pm$ 2.1 & 6.4 $\pm$ 1.5 & *47.5 $\pm$ 22.4 & *4.6 $\pm$ 2.2\\ \hline
J0119-SMG.10 & 01:19:09.5 & +02:08:00.2 & 0.90 & 7.1 $\pm$ 1.7 & 5.0 $\pm$ 1.2 & *1.8 $\pm$ 15.7 & *0.2 $\pm$ 1.5\\ \hline
J0119-SMG.11 & 01:18:52.5 & +02:06:36.2 & 0.93 & 9.0 $\pm$ 2.2 & 6.3 $\pm$ 1.5 & *34.7 $\pm$ 16.6 & *3.4 $\pm$ 1.6\\ \hline
J0119-SMG.12 & 01:19:07.9 & +02:09:44.2 & 0.95 & 8.6 $\pm$ 2.1 & 5.9 $\pm$ 1.4 & *32.4 $\pm$ 17.7 & *3.1 $\pm$ 1.7\\ \hline
J0119-SMG.13 & 01:19:24.3 & +02:05:56.2 & 0.96 & 7.0 $\pm$ 1.7 & 4.8 $\pm$ 1.2 & *27.3 $\pm$ 17.7 & *2.6 $\pm$ 1.7\\ \hline
J0119-SMG.14 & 01:19:15.5 & +02:02:46.2 & 0.97 & 6.1 $\pm$ 1.5 & 4.2 $\pm$ 1.0 & *-20.1 $\pm$ 18.3 & *-2.0 $\pm$ 1.8\\ \hline
J0119-SMG.15 & 01:18:50.8 & +02:06:14.2 & 0.99 & 8.1 $\pm$ 2.0 & 5.5 $\pm$ 1.4 & *0.9 $\pm$ 15.8 & *0.1 $\pm$ 1.5\\ \hline
J0119-SMG.16 & 01:19:17.6 & +02:06:02.2 & 0.99 & 6.2 $\pm$ 1.5 & 4.2 $\pm$ 1.0 & *12.9 $\pm$ 16.8 & *1.3 $\pm$ 1.6\\ \hline
\hline
\end{tabular}
\end{table*}

\begin{table*}
\caption{Position and fluxes of SMG candidates in the field of quasar pair J1622. For non-detections at 450~$\mu$m, we report forced fluxes at the coordinate of the source and mark them with an asterisk.}
\label{tab:smg1622}
\centering 
\begin{tabular}{c c c c c c c c}
\hline\hline
ID & RA (J2000) & Dec (J2000) & $\Delta(\alpha,\delta)$ [\arcsec] & $f_{850}$ [mJy] & $f_{\rm 850, db}$ [mJy] & $f_{450}$ [mJy] & $f_{\rm 450, db}$ [mJy]\\    
\hline
1622-SMG.1 & 16:22:07.7 & +07:01:51.5 & 0.79 & 5.6 $\pm$ 0.9 & 3.6 $\pm$ 0.6 & *13.0 $\pm$ 8.7 & *2.4 $\pm$ 1.6\\ \hline
1622-SMG.2 & 16:22:16.7 & +07:01:59.5 & 0.81 & 6.6 $\pm$ 1.2 & 4.1 $\pm$ 0.7 & *8.5 $\pm$ 10.6 & *1.6 $\pm$ 2.0\\ \hline
1622-SMG.3 & 16:21:49.1 & +07:02:23.5 & 0.82 & 9.3 $\pm$ 1.7 & 5.8 $\pm$ 1.0 & *6.9 $\pm$ 14.0 & *1.3 $\pm$ 2.6\\ \hline
1622-SMG.4 & 16:22:06.3 & +07:08:01.5 & 0.84 & 11.6 $\pm$ 2.2 & 6.9 $\pm$ 1.3 & *-2.2 $\pm$ 25.9 & *-0.4 $\pm$ 4.8\\ \hline
1622-SMG.5 & 16:21:52.9 & +07:00:17.5 & 0.85 & 7.6 $\pm$ 1.5 & 4.5 $\pm$ 0.9 & *8.0 $\pm$ 13.1 & *1.5 $\pm$ 2.4\\ \hline
1622-SMG.6 & 16:22:00.0 & +06:59:07.5 & 0.87 & 6.8 $\pm$ 1.4 & 3.9 $\pm$ 0.8 & *19.0 $\pm$ 15.2 & *3.5 $\pm$ 2.8\\ \hline
1622-SMG.7 & 16:22:17.6 & +07:00:29.5 & 0.87 & 6.4 $\pm$ 1.3 & 3.6 $\pm$ 0.7 & *12.7 $\pm$ 12.0 & *2.3 $\pm$ 2.2\\ \hline
1622-SMG.8 & 16:22:00.1 & +07:03:43.5 & 0.89 & 6.2 $\pm$ 1.3 & 3.4 $\pm$ 0.7 & *14.4 $\pm$ 12.8 & *2.6 $\pm$ 2.4\\ \hline
1622-SMG.9 & 16:22:20.0 & +06:57:45.5 & 0.89 & 7.6 $\pm$ 1.6 & 4.2 $\pm$ 0.9 & *6.6 $\pm$ 17.2 & *1.2 $\pm$ 3.2\\ \hline
1622-SMG.10 & 16:21:50.7 & +07:02:01.5 & 0.91 & 7.2 $\pm$ 1.6 & 3.9 $\pm$ 0.9 & *14.0 $\pm$ 13.0 & *2.6 $\pm$ 2.4\\ \hline
1622-SMG.11 & 16:22:15.9 & +06:56:51.5 & 0.92 & 8.4 $\pm$ 1.9 & 4.5 $\pm$ 1.0 & *-0.7 $\pm$ 21.8 & *-0.1 $\pm$ 4.0\\ \hline
1622-SMG.12 & 16:22:24.2 & +06:58:47.5 & 0.92 & 6.9 $\pm$ 1.6 & 3.6 $\pm$ 0.8 & *13.4 $\pm$ 15.8 & *2.5 $\pm$ 2.9\\ \hline
1622-SMG.13 & 16:22:32.8 & +07:00:49.5 & 0.93 & 8.4 $\pm$ 1.9 & 4.4 $\pm$ 1.0 & *7.3 $\pm$ 16.6 & *1.3 $\pm$ 3.0\\ \hline
1622-SMG.14 & 16:21:53.7 & +07:02:49.5 & 0.93 & 6.4 $\pm$ 1.5 & 3.3 $\pm$ 0.8 & *11.3 $\pm$ 13.2 & *2.1 $\pm$ 2.4\\ \hline
1622-SMG.15 & 16:22:16.9 & +06:58:05.5 & 0.96 & 5.8 $\pm$ 1.4 & 3.0 $\pm$ 0.7 & *-7.4 $\pm$ 14.7 & *-1.4 $\pm$ 2.7\\ \hline
1622-SMG.16 & 16:22:12.0 & +07:03:09.5 & 0.98 & 4.1 $\pm$ 1.0 & 2.1 $\pm$ 0.5 & *1.9 $\pm$ 9.2 & *0.3 $\pm$ 1.7\\ \hline
1622-SMG.17 & 16:22:00.4 & +07:05:13.5 & 0.98 & 6.1 $\pm$ 1.5 & 3.1 $\pm$ 0.7 & *1.8 $\pm$ 12.1 & *0.3 $\pm$ 2.2\\ \hline
1622-SMG.18 & 16:22:15.9 & +07:05:19.5 & 0.98 & 6.4 $\pm$ 1.5 & 3.2 $\pm$ 0.8 & *32.8 $\pm$ 13.7 & *6.0 $\pm$ 2.5\\ \hline
1622-SMG.19 & 16:22:10.5 & +06:57:57.5 & 0.99 & 6.5 $\pm$ 1.6 & 3.3 $\pm$ 0.8 & *-3.5 $\pm$ 14.6 & *-0.6 $\pm$ 2.7\\ \hline
1622-SMG.20 & 16:22:03.6 & +07:06:19.5 & 0.99 & 7.4 $\pm$ 1.8 & 3.7 $\pm$ 0.9 & *9.3 $\pm$ 14.3 & *1.7 $\pm$ 2.6\\ \hline
1622-SMG.21 & 16:22:21.5 & +06:59:03.5 & 1.0 & 5.6 $\pm$ 1.4 & 2.8 $\pm$ 0.7 & *-12.9 $\pm$ 14.1 & *-2.4 $\pm$ 2.6\\ \hline
1622-SMG.22 & 16:22:06.6 & +07:07:23.5 & 1.01 & 7.6 $\pm$ 1.9 & 3.8 $\pm$ 0.9 & *-14.0 $\pm$ 20.0 & *-2.6 $\pm$ 3.7\\ \hline
\hline
\end{tabular}
\end{table*}
\newpage\phantom{xyz}\newpage\phantom{xyz}\newpage\phantom{xyz}\newpage\phantom{xyz}\newpage\phantom{xyz}\newpage\phantom{xyz}\newpage\phantom{xyz}\newpage\phantom{xyz}\newpage\phantom{xyz}
\section{Testing the angle recovery algorithm in simulated filaments}
\label{app:testangle}
\begin{figure}
    \centering
    \includegraphics[width=\hsize]{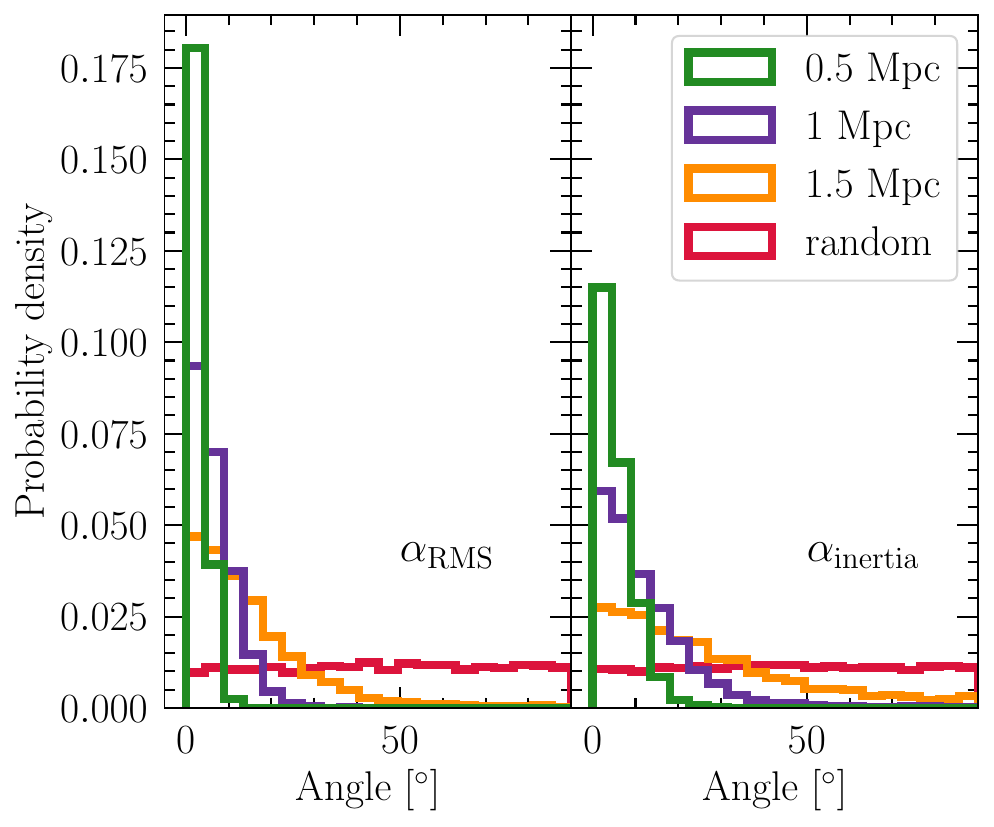}
    \caption{Absolute recovered angle $\alpha_{\rm RMS}$ (left panel) and $\alpha_{\rm inertia}$ (right panel) for the three tested filament widths 0.5~Mpc, 1~Mpc, and 1.5~Mpc, oriented at an angle of zero degrees, as well as random source distributions.}
    \label{fig:testanglehist}
\end{figure}

\begin{center}
\begin{minipage}{\hsize}
To test the success rate in recovering the angle of a filament as traced by SMGs, we created simulated maps of SMGs distributed within filaments of varying widths (0.5~Mpc, 1~Mpc, 1.5~Mpc and random distribution) and applied the algorithms described in Sec.~\ref{subsec:results:angle}. Specifically, for each quasar pair field and filament width, we randomly sampled fluxes between the minimum and maximum detected flux using the normalized source model (Table~\ref{tab:Schechter}) multiplied by the completeness (Fig.~\ref{fig:completeness}) as probability distribution. We then randomly chose a source position in the filament and only kept the simulated SMG if the position fell within the detectable area of the respective flux, until reaching the same source number as in the SMG candidate catalog for the field. Analogously to Sec.~\ref{subsec:results:angle}, we calculated $\alpha_{\rm rms}$ and $\alpha_{\rm inertia}$ in the resulting mock map and repeated the process 500 times for every filament width and the random source distributions.
The resulting angle probability densities for a stack of all fields is shown in Fig.~\ref{fig:testanglehist}.
While a two-sample Kolmogorov-Smirnov test between the different angles from filaments and the respective random distributions returns p-values below the data type precision, it becomes apparent that $\alpha_{\rm RMS}$ is more successful in recovering the injection angle of zero degrees: the angle distribution of $\alpha_{\rm RMS}$ for the thickest filament has a standard deviation of 20$^{\circ}$, while the standard deviation of the distribution for $\alpha_{\rm inertia}$ is 37$^{\circ}$.

\end{minipage}
\end{center}
\newpage

\end{appendix}

\end{document}